\documentclass[12pt,oupdraft]{sysbio}

\usepackage{indentfirst}
\usepackage{outlines}
\usepackage{fixltx2e}
\usepackage{textcomp}
\usepackage{amsfonts}
\usepackage{verbatim}
\usepackage[english]{babel}
\usepackage{pifont}
\usepackage{color}
\usepackage{setspace}
\usepackage{lscape}
\usepackage{indentfirst}
\usepackage[normalem]{ulem}
\usepackage{booktabs}
\usepackage{natbib}
\usepackage{float}
\usepackage{latexsym}
\usepackage{url}
\usepackage{graphicx}
\usepackage{amssymb}
\usepackage{amsmath}
\usepackage{bm}
\usepackage{array}
\usepackage{mhchem}
\usepackage{ifthen}
\usepackage{caption}
\usepackage{xcolor}
\usepackage{amsthm}
\usepackage{amstext}
\usepackage{subcaption}

\usepackage{amsmath, amsfonts, latexsym, amsthm, amssymb,mathrsfs,txfonts}
\usepackage{epsfig}

\newcounter{xxx}
\def\E{{\mathbb E}}        

\newcommand\X{{\cal X}}

\newcommand{\ud}{\,\mathrm{d}}

\newcommand{\xbold}{\boldsymbol{x}}

\newcommand{\rcess}{\text{rCESS}}
\newcommand{\varsmc}{\text{Var}_\text{SMC}}

\newcommand{\varbmc}{\text{Var}_\text{MC}}

\usepackage{algorithm, float}
\usepackage[noend]{algpseudocode}
\usepackage{setspace}

\renewcommand{\theequation}{\arabic{equation}}

\bibpunct{(}{)}{;}{a}{}{,}  

\begin{document}

\title{An Annealed Sequential Monte Carlo Method for Bayesian Phylogenetics}

 \author{Liangliang Wang$^{1,\ast}$, Shijia Wang$^{1,\ast}$, Alexandre Bouchard-C\^ot\'e$^{2,\ast}$\\[4pt]
 \textit{$^{1}$~Department of Statistics and Actuarial Science, Simon Fraser University, Burnaby, British Columbia, V5A 1S6, Canada}
 \\
 \textit{$^{2}$~Department of Statistics, University of British Columbia, Vancouver, British Columbia, V6T 1Z4, Canada}
 \\[2pt]
{\bf *Corresponding authors:} \textit{lwa68@sfu.ca / shijiaw@sfu.ca / bouchard@stat.ubc.ca}}

 \markboth%
 {Liangliang Wang, Shijia Wang and Alexandre Bouchard-C\^ot\'e}
 {An annealed SMC for Bayesian phylogenetics}

\maketitle

\begin{abstract}
{ We describe an ``embarrassingly parallel'' method for Bayesian phylogenetic inference, annealed Sequential Monte Carlo, based on recent advances in the Sequential Monte Carlo literature such as adaptive determination of annealing parameters. The algorithm provides an approximate posterior distribution over trees and evolutionary parameters as well as an unbiased estimator for the  marginal likelihood.
This unbiasedness property can be used for the purpose of testing the correctness of posterior simulation software. 
We evaluate the performance of phylogenetic annealed Sequential Monte Carlo by reviewing and comparing with other computational Bayesian phylogenetic methods, in particular, different marginal likelihood estimation methods.  Unlike previous Sequential Monte Carlo methods in phylogenetics, our annealed method can utilize standard Markov chain Monte Carlo tree moves and hence benefit from the large inventory of such moves available in the literature.  Consequently, the annealed Sequential Monte Carlo method should be relatively easy to incorporate into existing phylogenetic software packages based on Markov chain Monte Carlo algorithms.  We illustrate our method using simulation studies and real data analysis. }
{Sequential Monte Carlo; phylogenetics; {marginal likelihood}}
\end{abstract}

\section{Introduction} 

The Bayesian paradigm is widely used in systematic biology, principally for the purpose of  phylogenetic reconstruction as well as for evaluating the empirical support of evolutionary models \citep{Chen2014PhyloHandbook}. Both of  these tasks, 
Bayesian phylogenetic reconstruction and model selection, involve an intractable sum over topologies as well as a high dimensional integral over branch lengths and evolutionary parameters. 
Consequently, Markov chain Monte Carlo (MCMC) methods have been widely used in the past twenty years to approximate posterior distributions defined over the space of phylogenetic trees \citep{Rannala1996}. 
 
Despite their success, MCMC phylogenetic methods are still afflicted by two key limitations, hence motivating the need for alternative approximations method for posterior distributions over phylogenetic trees. 

 Firstly, MCMC methods do not readily take advantage of highly parallel computer architectures. This is problematic in the current context as progress in computational power mostly comes in the form of parallelism gains. 
While there are techniques available to parallelize phylogenetic MCMC methods, they are generally not ``embarrassingly parallel'': for example, parallel Metropolis coupled MCMC \citep{Altekar2004MCMCMC} may reach a point where the addition of cores actually reduces sampling efficiency \citep{atchade_towards_2011}.

A second challenge with MCMC-based phylogenetic approximations arises in the context of model selection. By comparing the  marginal likelihood $Z = p(y)$, where $y$ denotes observed data, under different models, one can approach scientific questions under the Bayesian framework while naturally taking into account differences in model complexity.  More specifically, the ratio $r = p_1(y)/p_2(y)$ of two marginal likelihoods based on two evolutionary models,  $p_1(\cdot), p_2(\cdot)$, can be used to assess the strength of evidence $y$ provides for $p_1$ (when $r > 1$) or $p_2$ (when $r < 1$).  The ratio $r$ is called the Bayes factor \citep{jeffreys1935some, lartillot_computing_2006, oaks2018marginal}. In the context of phylogenetics, the Bayes factor assesses how much support a set of sequencing data provides for one evolutionary model against another one. 

 Several methods have been proposed to estimate marginal likelihoods based on MCMC methods (\cite{newton_approximate_1994,gelman_simulating_1998,friel_marginal_2008}, \emph{inter alia}), including work tailored to the phylogenetic context  \citep{huelsenbeck_bayesian_2004,lartillot_computing_2006,xie2010improving,fan2010choosing}. 
However these methods all have different drawbacks, see for example the aptly named review, ``Nineteen dubious ways to compute the marginal likelihood of a phylogenetic tree topology'' \citep{fourment_19_2018}. 
Moreover, one limitation shared by all MCMC-based marginal likelihood estimators is that they are generally biased (in the technical sense of the term as used in  computational statistics, reviewed in the theory section of the paper)---unless one is able to initialize the MCMC chains to the exact stationary distribution, which in practice is not possible. We argue that in certain scenarios, it can be useful to have unbiased methods. One example we elaborate on is for the purpose of a new test to ascertain correctness of posterior simulation software. Another class of examples comes from the burgeoning field of pseudo-marginal methods \citep{andrieu_pseudo-marginal_2009}. 

Sequential Monte Carlo (SMC) methods (see \cite{Doucet11tutorial} for an accessible introduction to SMC) provide a flexible framework to construct unbiased estimators and past work has shown they can be very efficient in a phylogenetic context \citep{teh08a,gorur09,Bouchard2012Phylogenetic,Gorur2012ScalableSMC,LiangliangWang2015,everitt2016sequential,dinh2016online,smith2017infectious,fourment2017effective}. One drawback caused by the high degree of flexibility that comes with SMC is that the phylogenetic SMC algorithms developed so far are non-trivial to adapt to existing MCMC-based phylogenetic frameworks.   Here we propose a different construction based on the seminal work of \cite{del2006sequential}, in turn based on annealed importance sampling (AIS) \citep{neal_annealed_1998}, which yields an SMC method which is in a sense much closer to standard MCMC, while providing unbiased estimators of the marginal likelihood. The proposed method, which we call phylogenetic annealed SMC, can directly make use of any existing phylogenetic MCMC proposals, a rich literature covering many kinds of phylogenetic trees \citep{Rannala1996,yang1997bayesian,Mau1999,Larget1999,Li2000,Holder2003phylogeny,Rannala01082003,Lakner01022008,Hohna2008,Hohna2012}. It is easy to incorporate the proposed annealed SMC into existing phylogenetic software packages that implement MCMC algorithms, such as RevBayes \citep{Hohna2016} or BEAST \citep{Drummond2007}. At the same time, our method can leverage state-of-the-art advances in the field of adaption of SMC algorithms, making the algorithm fully automated in most cases.

Our implementation of the proposed method is available at \url{https://github.com/liangliangwangsfu/annealedSMC}. All our experimental setups and results are available at \url{https://github.com/shijiaw/AnnealingSimulation}. The algorithms described here are also available in the Blang probabilistic programming language \url{https://github.com/UBC-Stat-ML/blangSDK}, which supports a small but growing set of phylogenetic models.

\section{Literature review}
There is a growing body of work on SMC-based Bayesian phylogenetic inference. Indeed, a powerful feature of the general SMC framework \citep{del2006sequential} is that the space on which the distributions $\pi_r$ are defined is allowed to vary from one iteration to the next. All previous work on SMC methods for phylogenetics has exploited this feature for various purposes reviewed here. 

In one direction, several ``bottom up'' approaches  \citep{teh08a,gorur09,Bouchard2012Phylogenetic,Gorur2012ScalableSMC,LiangliangWang2015} have been proposed to allow more efficient reuse of intermediate stages of the Felsenstein pruning recursions. For these methods, the intermediate distributions are defined over forests over the observed taxa, and hence their dimensionality increases with $r$. These methods are most effective in clock or nearly-clock trees. For general trees, it is typically necessary to perform additional MCMC steps, which makes it harder to use in the context of estimation of marginal likelihoods.

In a related direction, \cite{dinh2016online} and \cite{fourment2017effective} use a sequence of targets where $\pi_r$ is a tree over the first $r$ tips. This construction is especially useful in scenarios where taxonomic data  come in an online fashion. 

Another use case of SMC methods in phylogenetics arises from Bayesian analysis of intractable evolutionary models. For example, SMC has been used for Bayesian phylogenetic analysis based on infinite state-space evolutionary models \citep{Hajiaghayi2014Efficient} or for joint inference of transmission networks \citep{smith2017infectious}.

Finally, a concurrent line of work \citep{everitt2016sequential}  has explored a combination of reversible jump methods with phylogenetic models.

One drawback of letting the dimensionality of $\pi_r$ vary with $r$ as all the above methods do, is that it makes it significantly harder to incorporate SMC into existing Bayesian phylogenetic inference packages such as MrBayes \citep{huelsenbeck_mrbayes:_2001}, RevBayes or BEAST. In contrast, in our method the target distributions $\pi_r$ are all defined over the same space. The annealed SMC framework in this context utilizes Metropolis-Hastings kernels in the inner loop but combines them in a different fashion compared to standard MCMC algorithms, or even compared to parallel tempering MCMC algorithms.

\section{Setup and notation}

We let $t$ denote a phylogenetic tree with tips labelled by a fixed set of operational taxonomic units $X$. The variable $t$ encapsulates the tree topology and a set of positive branch lengths. Our methodology is directly applicable to any class of phylogenetic trees where MCMC proposal distributions are available. This includes for example clock trees \citep{Hohna2008} as well as non-clock trees \citep{Lakner01022008}. 

We let $\theta$ denote evolutionary parameters, for example the parameters of a family of rate matrices such as the general time reversible (GTR) model  \citep{tavare1986some}, or diffusion parameters in the case of continuous traits \citep{lemey2010phylogeography}. Again our method is applicable to any situation where MCMC proposals are available for exploring the space of $\theta$.
We use $x = (t, \theta)$ to denote these two latent variables.

We let $y$ denote observed data indexed by the tips $X$ of $t$. We assume a likelihood function $p(y | x)$ is specified such that for any hypothesized tree and parameters, the value $p(y | x)$ can be computed efficiently. This assumption is sometimes called pointwise computation. This is a typical assumption in Bayesian phylogenetics, where this computation is done with some version of Felsenstein pruning \citep{felsenstein1973maximum, felsenstein1981} (an instance of the Forward-Backward algorithm \citep{forney1973viterbi}).

Finally, let $p(x)$ denote a prior on the parameters and trees, which we assume can also be computed pointwise efficiently. This defines a joint distribution, denoted $\gamma(x) = p(x) p(y | x)$. We ignore the argument $y$ from now on since the data is viewed as fixed in a Bayesian analysis context. 

We are interested in approximating a posterior distribution on $x$ given data $y$, denoted:
\begin{align}\label{eq:posterior}
\pi(x) = \frac{\gamma(x)}{\int \gamma(x') \ud x'}.
\end{align}
Here the integral and $\ud x'$ are viewed in an abstract sense and include both summation over discrete latent variables such as topologies and standard integration over continuous spaces. 

The denominator can be interpreted as the marginal likelihood under the model specified by the prior and likelihood functions, which we denote by $Z$:
\begin{align}\label{eq:data-like}
Z = p(y) = \int \gamma(x) \ud x.
\end{align}
Computation of this quantity, also called the normalization constant or evidence, is the main challenge involved when doing Bayesian model selection. 

Other quantities of interest include expectations with respect to the posterior distribution, characterized by a real-valued function of interest $f$ based on which we would like to compute
\begin{equation}\label{eq:expectation} 
\int \pi(x) f(x) \ud x.
\end{equation}
For example if we seek a posterior clade support for a subset $X' \subset X$ of the leaves $X$, 
\[ f(x) = f(t,\theta) = 1[  \textrm{$t$ admits $X'$ as a clade}], \]
where $1[s]$ denotes the indicator function which is equal to one if the boolean expression $s$ is true and zero otherwise.

\section{Annealed SMC for phylogenetics}

\subsection{Sequences of Distributions}

In standard MCMC methods, we are interested in a single probability distribution, the posterior distribution. 
However, there are several reasons why we may use a \emph{sequence} of distributions rather than only one. 

A first possibility is that we may have an online problem, where the data is revealed sequentially and we want to perform inference sequentially in time based on the data available so far. The distribution at step $r$ is then the posterior distribution conditioning on the first $r$ batches of data. This approach is explored in the context of phylogenetics in \cite{dinh2016online}, where a batch of data consists in genomic information for one additional operational taxonomic unit. We do not pursue this direction here but discuss some possibilities for combinations in the discussion. 

A second reason for having multiple distributions, and the focus of this work, is to facilitate the exploration of the state space.  This is achieved for example by raising the likelihood term to a power $\phi_r$ between zero and one, which we multiply with the prior
\begin{align}\label{eqn:piTophi}
\gamma_r(x) = p(y|x)^{\phi_r} p(x).
\end{align}
 MCMC may get stuck in a region of the space of phylogenetic trees around the initial value. This may happen for example around a local maximum (mode) in the posterior density. Such a region is sometimes called a ``basin of attraction'', and no single basin of attraction may be enough to well represent the full posterior distribution.  Introducing a series of powered posterior distributions can alleviate this issue.  A small value of $\phi_r$ flattens the posterior and makes MCMC samplers move easily between the different basins of attractions. The samples are initially overly dispersed but are then coerced into the posterior distribution $\pi(x)$ by slowly increasing the annealing parameter $\phi_r$.  
 
We do not anneal the prior to ensure that $\gamma_r(x)$ has a finite normalization constant,
\begin{align*}
\int \gamma_r(x) \ud x &= \E_{p(x)}[(p(y|X))^{\phi_r}] \\
&\le \left( \E_{p(x)} [p(y|X)] \right)^{\phi_r} = \left( p(y) \right)^{\phi_r} < \infty,
\end{align*}
where the first inequality follows from the concavity of $(\cdot)^{\phi_r}$ and Jensen's inequality.

A third scenario is that we may encounter a ``tall data'' problem, e.g. biological sequences with a large number of sites. When the number of sites is large, evaluation of the unnormalized posterior $\gamma_{r}(x)$ defined in Equation (\ref{eqn:piTophi})
 is computationally expensive. The idea of data subsampling \citep{quiroz2018speeding, quiroz2018speeding2, bardenet2017markov, gunawan2018subsampling} could be used to define the sequence of distributions. The construction of the sequence of  distributions is described in Appendix 1.  

The probability distributions 
\begin{align}
\pi_r(x) = \frac{\gamma_r(x)}{\int \gamma_r(x') \ud x'}
\end{align}
are therefore well defined and we denote their respective  normalization constants by
\begin{align}
Z_r = \int \gamma_r(x) \ud x.
\end{align}

If the exponent $\phi_r$ is zero, then the distribution $\pi_r$ becomes the prior which is often easy to explore and in fact independent samples can be extracted in many situations. At the other extreme, the distribution at power $\phi_r = 1$ is the distribution of interest. 

The intermediate distributions $\{\pi_r \}_{r=1,\ldots, R}$ are defined on a common measurable space $(\X, \mathcal{E})$.  The annealed SMC is a generalization of the standard SMC method \citep{Doucet01}. In standard SMC, the
intermediate distributions are defined on a space of strictly increasing dimension. 

\subsection{Basic Annealed SMC Algorithm}\label{sec:basic-algo}

We now turn to the description of annealed SMC in the context of Bayesian phylogenetic inference. The algorithm fits into the generic framework of \emph{SMC samplers} \citep{del2006sequential}: at each iteration, indexed by $r = 1, 2, \dots, R$, we maintain a collection indexed by $k \in \{1, 2, \dots, K\}$ of imputed latent states $x_{r,k}$, each paired with a non-negative number called a weight $w_{r,k}$; such a pair is called a particle. A latent state in our context consists in a hypothesized tree $t_{r,k}$ and a set of evolutionary parameters $\theta_{r,k}$, i.e. $x_{r,k} = (t_{r,k}, \theta_{r,k})$. In contrast to previous SMC methods, $x_{r,k}$ is always of the same data type: no partial states such as forest or trees over subsets of leaves are considered here. 

A particle population consists in a list of particles $(x_{r,\cdot}, w_{r,\cdot}) = \{(x_{r,k}, w_{r,k}) : k\in\{1, \dots, K\}\}$. A particle population can be used to estimate posterior probabilities as follows: first, normalize the weights, denoted after normalization using capital letter, $W_{r,k} = w_{r,k} / \sum_{k'} w_{r,k'}$. Second, use the approximation:
\begin{equation}\label{eq:expectation-approx} 
\int \pi_r(x) f(x) \ud x \approx \sum_{k=1}^K W_{r,k} f(x_{r,k}).
\end{equation}
For example if we seek a posterior clade support for a subset $X' \subset X$ of the leaves $X$, this becomes
\[ \sum_{k=1}^K W_{r,k} 1[  \textrm{sampled tree $t_{r,k}$ admits $X'$ as a clade}]. \]
The above formula is most useful at the last SMC iteration, $r = R$, since $\pi_R$ coincides with the posterior distribution by construction.

At the first iteration, each of the particles' tree and evolutionary parameters are sampled independently and identically from their prior distributions. We assume for simplicity that this prior sampling step is tractable, a reasonable assumption in many phylogenetic models.  
After initialization, we therefore have a particle-based approximation of the prior distribution. Intuitively, the goal behind the annealed SMC algorithm is to progressively transform this prior distribution approximation into a posterior distribution approximation. 

To formalize this intuition, we use the sequence of distributions introduced in the previous section. The last ingredient required to construct an SMC algorithm is an SMC proposal distribution ${K}_r(x_{r-1,k}, x_{r,k})$, used to sample a particle for the next iteration given a particle from the previous iteration. Since $x_{r-1,k}$ and $x_{r,k}$ have the same dimensionality in our setup, it is tempting to use MCMC proposals $q_r(x_{r-1,k}, x_{r,k})$ in order to build SMC proposals, for example, subtree prune and regraft moves, and Gaussian proposals for the continuous parameters and branch lengths. 
Indeed, there are several advantages of using MCMC proposals as the basis of SMC proposals. First, this means we can leverage a rich literature on the topic \citep{Rannala1996,yang1997bayesian,Mau1999,Larget1999,Li2000,Holder2003phylogeny,Rannala01082003,Lakner01022008,Hohna2008,Hohna2012}. Second, it makes it easier to add SMC support to existing MCMC-based software libraries. Third, it makes certain benchmark comparison between SMC and MCMC more direct, as we can then choose the set of moves to be the same for both. On the flip side, constructing MCMC proposals is somewhat more constrained, so some of the flexibility provided by the general SMC framework is lost.

Naively, we could pick the SMC proposal directly from an MCMC proposal, ${K}_r(x_{r-1,k}, x_{r,k}) = q_r(x_{r-1,k}, x_{r,k})$. However, doing so would have the undesirable property that the magnitude of the fluctuation of the weights of the particles from one iteration to the next, $\|W_{r-1,\cdot} - W_{r,\cdot}\|$, does not converge to zero when the annealing parameter change $\phi_r - \phi_{r-1}$ goes to zero. This lack of convergence to zero can potentially cause severe particle degeneracy problems, forcing the use of a number of particles larger than what can be realistically accommodated in memory (although workarounds exist, e.g. \cite{Jun2014Memory}). 
 To avoid this issue, we follow \cite{del2006sequential} and use as SMC proposal the accept-reject Metropolis-Hastings transition probability based on $q_r$ (called a Metropolized proposal), reviewed in Algorithm \ref{algo:MHalgorithm}.

\begin{algorithm}
   \caption{\bf{Accept-reject Metropolis-Hastings algorithm  }}
  \label{algo:MHalgorithm}
   {\fontsize{12pt}{12pt}\selectfont
\begin{algorithmic}[1]
\State Propose a new tree and/or new evolutionary parameters, $x_r^*\sim q_r(x_{r-1},\cdot)$.
 \Comment   For example, using a nearest neighbour interchange, and/or a symmetric normal proposal on branch lengths and/or evolutionary parameters.
  
\State 	Compute the Metropolis-Hastings ratio based on $\gamma_r$:
 	\begin{eqnarray*}	
 		\alpha_r(x_{r-1}, x_r^*)=\min \left\{1, \frac{\gamma_r(x_r^*)q(x_{r}^*,
 			x_{r-1})}{\gamma_r(x_{r-1})q(x_{r-1}, x_r^*)}\right\}.
 	\end{eqnarray*}
  
   \State Simulate $u\sim U(0,1)$.
    
\If{$u < \alpha_r(x_{r-1}, x_r^*)$}
\State  $x_r = x_r^*$.   \Comment{Output the proposal $x_r^*$.}

  \Else
    \State $x_r = x_{r-1}$. \Comment{Output the previous state $x_{r-1}$.}
  \EndIf

\end{algorithmic}
}
\end{algorithm}

The key point is that a theoretical argument (reviewed in the Appendix 2) shows that provided that (1) ${K}_r$ has stationary distribution $\pi_r$ (which is true by construction,  a consequence of using the Metropolis-Hastings algorithm) and (2) we use the weight formula:
\begin{equation}\label{eq:weight-update}
w_{r,k} = \frac{\gamma_r}{\gamma_{r-1}}(x_{r-1, k}),
\end{equation}
then we obtain a valid SMC algorithm, meaning that the key theoretical properties expected from SMC hold under regularity conditions, see Section \emph{Theoretical Properties}. 

In the important special case where $\gamma_r(x_r)$ is equal to the prior times an annealed likelihood, we obtain
\begin{equation}\label{eq:weight-update-detail}
w_{r,k} = [p(y|x_{r-1,k})]^{\phi_r - \phi_{r-1}}.
\end{equation}

As hoped, the update shown in Equation~(\ref{eq:weight-update-detail}) has the property that weight fluctuations vanish as the annealing parameter difference $\phi_r - \phi_{r-1}$ goes to zero. This will form the basis of the annealing parameter sequence adaptation strategies described in the next section. 
But for now, assume for simplicity that the number of iterations $R$ and the annealing schedule $\phi_r$, $r\in \{1, \dots, R\}$ is pre-specified. For example, a simple choice for the annealing parameter sequence  \citep{friel_marginal_2008} is  $\phi_r=(r/R)^{3}$, where $R$ is the total number of SMC iterations. 
In this case, the difference between successive annealing parameters is $(3r^{2}-3r+1)/R^{3}$.  An annealed SMC with  a larger value of
$R$  is computationally more expensive but has a  better performance.   

In contrast to other SMC algorithms, the annealed SMC algorithm does not require pointwise evaluation of the proposal ${K}_r(x_{r-1,k}, x_{r,k})$, i.e. given $x_{r-1,k}$ and a sampled $x_{r,k}$, we do not need to compute the numerical value of ${K}_r(x_{r-1,k}, x_{r,k})$ as it does not appear in the weight update formula, Equation~(\ref{eq:weight-update}). This point is important, since for Metropolis-Hastings kernels, pointwise evaluation would require computation of a typically intractable integral under the proposal in order to compute the total probability of rejection. The theoretical justification as to why we do not need pointwise evaluation of ${K_{r}}$ is detailed in Appendix 2.

In practice, many proposals are needed to modify different latent variables and to improve mixing. We give in  Appendix 3 the list of MCMC proposals we consider. Let $q_r^i$, $i=1,\ldots,M$, denote the various proposals, and ${K}_r^i$ the corresponding Metropolized transition probabilities. We need to combine them into one proposal ${K}_r$. To ensure that condition (1) above is satisfied, namely that ${K}_r$ obeys global balance with respect to $\pi_r$, use the following property \citep{Tierney94markovchains,Andrieu2003}: if each of the transition kernels
$\{{K}^i\},i=1,\ldots,M$, respects global balance with respect to $\pi$, then the \emph{cycle
	hybrid kernel} $\prod_{i=1}^M {K}^i$ and the \emph{mixture hybrid kernel} 
$\sum_{i=1}^Mp_i {K}^i, \sum_{i=1}^Mp_i=1$, also satisfy global balance with respect to $\pi$. The global balance condition, $\int \pi_r(x) {K}_r(x, x') \ud x = \pi_r(x')$, ensures that the Markov chain encoded by $K_r$ admits $\pi_r$ as a stationary distribution.  In practice, the mixture kernel is implemented by randomly selecting  $K^i$ with probability $p_{i}$ at each iteration \citep{Andrieu2003}. 

\begin{figure}
	\includegraphics[width=1\textwidth]{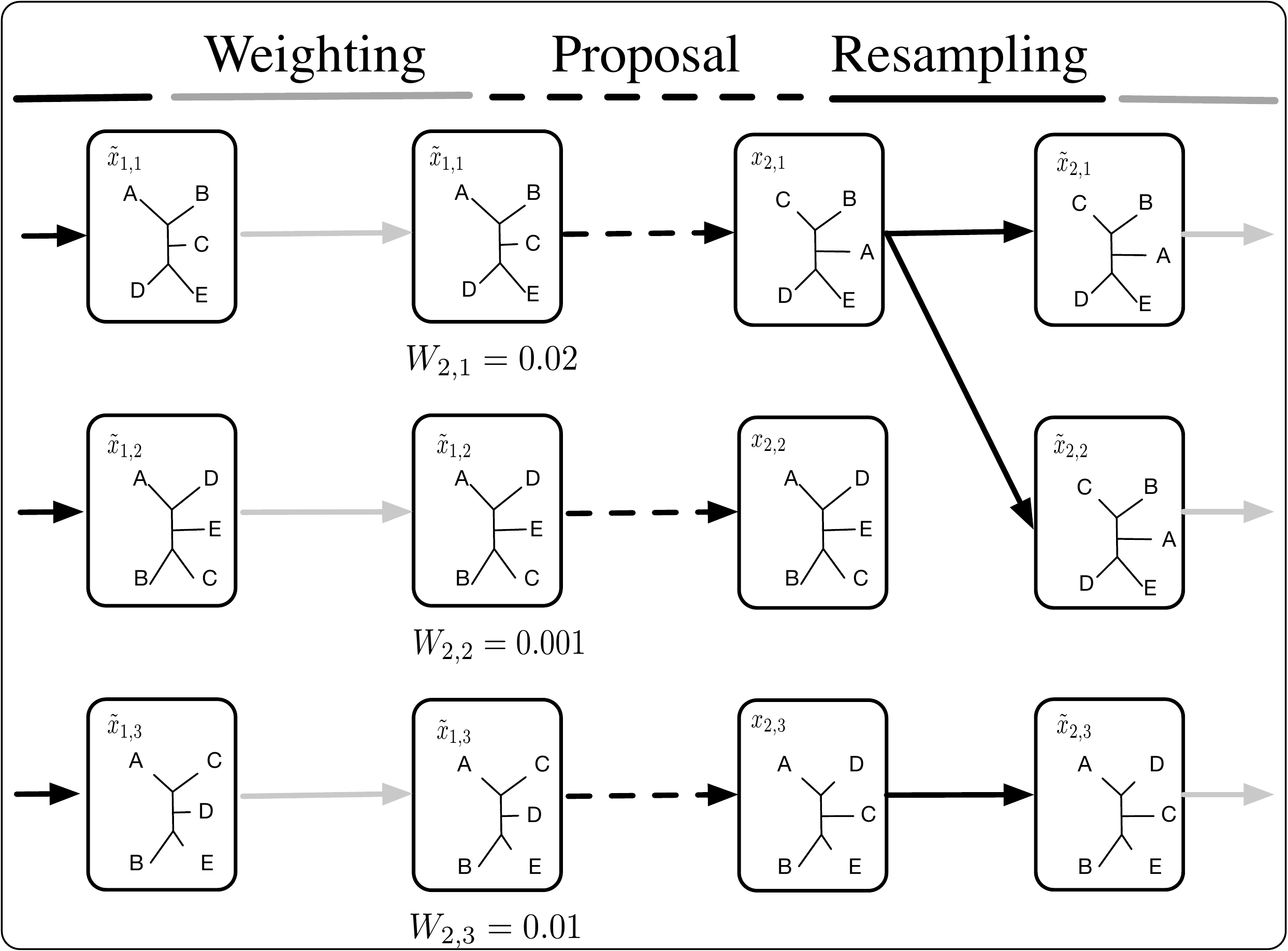}
	\caption{ An overview of the annealed SMC algorithmic framework for phylogenetic trees. The algorithm iterates the following three steps: (i) compute the weights using samples from the previous iteration, (ii) perform MCMC moves to propose new samples, and
		(iii) resample from the weighted samples to obtain an unweighted set of samples.}
	\label{fig: illustrationofSMC}
\end{figure}

We can now introduce in Algorithm  \ref{algo:simple-smc} the simplest version of the annealed SMC, which alternates between reweighting, propagating, and resampling. Figure \ref{fig: illustrationofSMC} presents an overview of the annealed SMC algorithmic framework. In the proposal step, we propose new particles through MCMC moves (typically Metropolis-Hastings moves). Finally, we use resampling to prune particles with smalls weights. A list of unweighted particles is obtained after the resampling step.  

 In the annealed SMC algorithm, note that the weighting and proposal steps can be interchanged. This is different from standard SMC algorithms, where in general the proposal has to be computed before weighting. This interchange is possible  because in the annealed SMC algorithm, the weighting function, Equation (\ref{eq:weight-update}), only depends on particles from the previous iteration and not from those just proposed as in standard SMC algorithms. This flexibility will come handy when designing adaptive schemes. 

\begin{algorithm}
   \caption{\bf{The simplest version of annealed SMC algorithm (for pedagogy)   }}
  \label{algo:simple-smc}
  {\fontsize{12pt}{12pt}\selectfont
\begin{algorithmic}[1]
  \State {\bfseries Inputs:} 
  \State (a) Prior over evolutionary parameters and trees, $p(x)$, where $x = (\theta, t)$; 
  \State (b) Likelihood function $p(y | x)$; 
  \State (c) Sequence of annealing parameters $0 = \phi_0 < \phi_1 < \dots < \phi_R = 1$.

   \State {\bfseries Outputs:}  Approximation of the posterior distribution, $\sum_k  \tilde W_{R,k} \delta_{\tilde x_{R,k}}(\cdot) \approx \pi(\cdot)$.

	\State Initialize SMC iteration index: $r \leftarrow 0$. 

	\State  Initialize annealing parameter: $\phi_r \leftarrow 0$.
   \For{$k \in \{1, 2, \dots, K\}$ } 
    \State  Initialize particles $x_{0, k} \leftarrow (\theta_{0,k}, t_{0,k}) \sim p(\cdot)$. 
	\State  Initialize weights to unity: $w_{0,k} \leftarrow 1$. 
\EndFor    
 \For{$r \in\{ 1, 2, \dots R\}$}
    \For{$k \in \{1, 2, \dots, K\}$}
  		\State \label{step:simple-prop} Sample particles $\tilde x_{r,k} \sim {K}_r(x_{r-1,k}, \cdot)$; ${K}_r$ is a $\pi_r$-invariant Metropolis-Hastings kernel.
		\State \label{step:simple-weigh} Compute unnormalized weights: $w_{r,k} =  [p(y|x_{r-1,k})]^{\phi_r - \phi_{r-1}}$.
\EndFor
\If{$r < R$}
\For{$k \in \{1, 2, \dots, K\}$}
    \State Resample the particles: $x_{r,k} \sim \sum_{k'}  \tilde W_{r,k'} \delta_{\tilde x_{r,k'}}(\cdot)$.
    \EndFor
  \Else
    \State No resampling needed at the last iteration.
  \EndIf
  
\EndFor 
\State Return the particle population $\tilde x_{r, \cdot}, \tilde W_{r,\cdot}$. 
\end{algorithmic}
}
\end{algorithm}

Before moving on to more advanced versions of the algorithm, we provide first some intuition to motivate the need for resampling.  
Theoretically, the algorithm produces samples from an artificial distribution with state space $\X \times \X \times \dots \times \X = \X^R$ (this is described in more detail in Appendix 2). However since we only make use of one copy of $\X$ (corresponding to the particles at the final SMC iteration), we would like to decrease the variance of  the state at iteration $R$ (more precisely, of Monte Carlo estimators of functions of the state at iteration $R$). This is what resampling for iteration $r < R$ achieves, at the cost of increasing the variance for the auxiliary part of the state space $r < R$. From this argument, it follows that resampling at the last iteration should be avoided. 

When resampling is performed at every iteration but the last, an estimate of the marginal likelihood, $p(y)$, is given by the product of the average unnormalized weights, namely:
\begin{equation}\label{eq:basic-z-estimator}
\hat Z_K := \prod_{r=1}^R \frac{1}{K} \sum_{k=1}^K w_{r,k}.
\end{equation}

\section{Adaptive mechanisms for annealed SMC}\label{sec:adapt}

We discuss how two adaptive schemes from the SMC literature can be applied in our Bayesian phylogenetic inference setup to improve the scalability and usability of the algorithm described in the previous section.  The first scheme relaxes the assumption that resampling is performed at every step, and the second is a method for automatic construction of the annealing parameter sequence. The two mechanisms go hand in hand and we recommend using both simultaneously. The combination yields Algorithm \ref{algo:adapt} which we explain in detail in the next two subsections. 

The two adaptive mechanisms make theoretical analysis considerably more difficult. This is a common situation in the SMC literature. A common work-around used in the SMC literature is to run the algorithm twice, a first time to adaptively determine the resampling and annealing schedules, and then a second independent time using the schedule fixed in the first pass. We call it debiased adaptive annealed SMC. 

\begin{algorithm}
   \caption{\bf{An adaptive annealed SMC algorithm   }}
  \label{algo:adapt}
{\fontsize{12pt}{12pt}\selectfont
\begin{algorithmic}[1]
  \State {\bfseries Inputs:} (a) Prior over evolutionary parameters and trees $p(x)$, where $x = (\theta, t)$;  
(b) Likelihood function $p(y | x)$.

   \State {\bfseries Outputs:}  (a) Approximation $Z$ of the marginal data likelihood, $Z \approx p(y) = \int p(\text{d} x) p(y | x)$; (b) Approximation of the posterior distribution, $\sum_k \tilde W_{R,k} \delta_{\tilde x_{R,k}}(\cdot) \approx \pi(\cdot)$.

	\State  Initialize SMC iteration index: $r \leftarrow 0$.  

	 \State  Initialize annealing parameter: $\phi_r \leftarrow 0$. 
	\State  Initialize marginal likelihood estimate: $Z\gets 1$. 
   \For{$k \in \{1, 2, \dots, K\}$}
    \State  Initialize particles with independent samples:  $x_{0, k} \leftarrow (\theta_{0,k}, t_{0,k}) \sim p(\cdot)$. 
	\State  Initialize weights to unity: $w_{0,k} \leftarrow 1$. 
\EndFor    
 \For{$r\in \{1, 2, \dots\}$}
    \State  Determine next annealing parameter: $\phi_r = \text{NextAnnealingParameter}(x_{r-1,\cdot}, w_{r-1,\cdot}, \phi_{r-1})$. 
    \For{$k \in \{1, \dots, K\}$}
      \State \label{step:adapt-weigh}  Compute pre-resampling unnormalized weights: $\tilde w_{r,k} = w_{r-1,k} [p(y|x_{r-1,k})]^{\phi_r - \phi_{r-1}}$. 
       \State Sample particles $\tilde x_{r,k} \sim {K}_r(x_{r-1,k}, \cdot)$; ${K}_r$ is a $\pi_r$-invariant Metropolis-Hastings kernel.
    \EndFor    
     \If{$\phi_r = 1$} 
   \State \label{step:adapt-return} update  $Z \gets (Z/K)\cdot \sum_k \tilde w_{r,k}$, then return updated $Z$ and particle population $\tilde x_{r, \cdot}, \tilde W_{r,\cdot}$.
      \Else
 
  \If{particle degeneracy is too severe, i.e. $\text{rESS}(\tilde W_{r,\cdot}) < \epsilon$}  
	\State \label{step:adapt-update-z} Update marginal likelihood estimate,  $Z \gets (Z/K)\cdot \sum_k \tilde w_{r,k}$. 
	\State Resample the particles. 
    \For{$k \in \{1, \dots, K\}$}
    \State \label{step:adapt-reset-weights} Reset particle weights:  $w_{r,k} = 1$.
    \EndFor
    \Else
    \For{$k \in \{1, \dots, K\}$}
    \State $w_{r,k} = \tilde w_{r,k}$; $\ x_{r,k} = \tilde x_{r,k}$. \Comment{ No resampling is needed.}
    \EndFor
   \EndIf
     \EndIf
\EndFor 
 
\end{algorithmic}
}
\end{algorithm}

\subsection{Measuring Particle Degeneracy using Relative (Conditional) Effective Sample Size (ESS)}

Both adaptive methods rely on being able to assess the quality of a particle approximation. For completeness, we provide more background in Appendix~5 on the notions of Effective Sample Size (ESS) and conditional ESS (CESS), a recent generalization which we use here \citep{ZHOU2016TowardAutomaticModelComparison}.  The notion of ESS in the context of importance sampling (IS) or SMC is distinct from the notion of ESS in the context of MCMC. The two are related in the sense of expressing a variance inflation compared to an idealized Monte Carlo scheme but they differ in the details. We will assume from now on that ESS refers to the SMC context.

We will use a slight variation of the definition of ESS and CESS where the measures obtained are normalized to be between zero and one. Some tuning parameters of the adaptive algorithms are easier to express in this fashion.  We use the terminology relative (conditional) ESS to avoid confusion.  Motivated by the analysis of the error of Monte Carlo estimators,  the key measure of particle degeneracy needed in the following is the relative conditional effective sample size:
\begin{align}\label{eq:rcess-def}
\text{rCESS}(W, u) =  
\left( 
\sum_{k=1}^K W_{k} u_k  
\right)^2 
\Big\slash
\sum_{k=1}^K W_{k} 
u_k^2,
\end{align}
where $W = (W_1, W_2, \dots, W_K)$ is a vector of weights of a set of reference weighted particles being updated using a vector of non-negative values $u = (u_1, u_2, \dots, u_K)$.  What $W$ and $u$ specifically represent will be explained in the next subsection. 

Having a high rCESS value is a necessary but not sufficient condition for a good SMC approximation. If it is low during some of the intermediate SMC iterations, then the ESS at the final iteration may not be representative of the true posterior approximation quality.

\subsection{Dynamic Resampling}\label{sec:adapt-resampling}

As explained in Section {\it Basic Annealed SMC Algorithm}\ref{sec:basic-algo}, the construction of the proposal guarantees that as the difference $\phi_r - \phi_{r-1}$ goes to zero, the fluctuation of the weights vanishes. In this context (of having small weight updates), resampling at every iteration is wasteful. Fortunately, SMC algorithms can be modified to forgo a subset of the resampling steps. From a theoretical stand-point, this is achieved by ``grouping'' the SMC proposals when they are not separated by a resampling round (and grouping similarly the intermediate distributions $\gamma_r$). For example, to resample every other round, use a transformed SMC algorithm with proposal ${K}'_{r/2}(x_r, (x_{r+1}, x_{r+2})) = {K}_{r+1}(x_r, x_{r+1}) {K}_{r+2}(x_{r+1}, x_{r+2})$, for each even $r$. For convenience, this can be implemented as an algorithm over $R$ iterations instead of $R/2$, with two modifications: first, when resampling is skipped, we multiply the weights; otherwise, we reset the weights to one after resampling. This is implemented in Lines~\ref{step:adapt-weigh} and \ref{step:adapt-reset-weights} of Algorithm~\ref{algo:adapt}. Second, we only use the weights corresponding to resampling rounds in the estimate of the marginal likelihood (Equation~(\ref{eq:basic-z-estimator})). This is implemented in Lines~\ref{step:adapt-return} and  \ref{step:adapt-update-z} of Algorithm~\ref{algo:adapt}.

Instead of specifying in advance the subset of iterations in which resampling should be performed, it is customary in the SMC literature to determine whether to resample in an adaptive fashion \citep{Doucet11tutorial}. To do so, the standard approach is to compute a measure of particle degeneracy at every iteration, and to perform resampling only when the particle degeneracy exceeds a pre-determined threshold. In Appendix 4, we empirically compare the performance of adaptive annealed SMC algorithm with different resampling thresholds. All our numerical experiments use the multinomial resampling method, but we recommend more advanced schemes such as stratified resampling \citep{douc2005comparison}.

The standard measure of particle degeneracy used for this purpose is called the relative ESS, defined as:

\begin{equation}\label{eq:simple-ess}
\text{rESS}(\tilde W_{r,\cdot}) = \left(K \sum_{k=1}^K \tilde W_{r,k}^2\right)^{-1}.
\end{equation}

The above formula can be shown to be a special case  of  rCESS, Equation~(\ref{eq:rcess-def}), as follows. Let $r^*$ denote the iteration of the latest resampling round preceding the current iteration $r$. This implies $W_{r^*, k} = 1/K$ for all $k$. Plugging in the weight update $u_{k} = \tilde w_{r,k}$ into Equation~(\ref{eq:rcess-def}), we obtain
\begin{align*}
\rcess(W_{r^*, \cdot}, \tilde w_{r, \cdot}) &= 
	\left( 
	\sum_{k=1}^K \frac{1}{K} \tilde w_{r, k} 
	\right)^2 
	\Big\slash
	\sum_{k=1}^K \frac{1}{K} 
	\tilde w_{r, k}^2 \\
	&= \frac{1}{K} \frac{\left( \sum_{k=1}^K  \tilde w_{r,k} \right)^2}{\sum_{k=1}^K  \tilde w_{r,k}^2} \\
	&= \left(K \sum_{k=1}^K \tilde W_{r,k}^2\right)^{-1}.
\end{align*}

\subsection{Adaptive Determination of Annealing Parameters}\label{sec:adapt-anneal}

Our sequence of intermediate artificial distributions $\pi_r$ as defined in Equation (\ref{eqn:piTophi}) is determined by the choice of the annealing schedule, $\{\phi_r\}$, or equivalently, by choosing the successive differences $\phi_r - \phi_{r-1}$. Ideally, the sequence of  intermediate distributions changes gradually from the prior distribution ($\phi_0 = 0$) to the posterior distribution ($\phi_R = 1$) so that the propagated particles from the current iteration can well approximate the next intermediate distribution.

In practice constructing such a sequence $\{\phi_r\}_{r=1, \ldots, R}$ is difficult and inconvenient. Not only the number of distributions $R$ to get a certain accuracy may depend on the number of taxa, the number of sites, and the complexity of the evolutionary model, but also the optimal spacing between consecutive annealing parameters is in general non-regular. To alleviate this, in the following we borrow an adaptive strategy from the Approximate Bayesian Computation literature \citep{del_moral_adaptive_2012}, also generalized to Bayesian model selection in \cite{ZHOU2016TowardAutomaticModelComparison}. 

The adaptive annealing scheme is based on two observations. First, our discrete set of intermediate distributions $\pi_1, \pi_2, \dots, \pi_R$ are actually continuously embedded into a continuum of distributions indexed by $\phi\in [0,1]$. Second, in the SMC algorithm presented in Algorithm \ref{algo:simple-smc}, the weight update, Line~\ref{step:simple-weigh}, depends only on $x_{r-1}$ (whereas in general SMC algorithms, the weight update could depend on both $x_{r-1}$ and $x_r$; here it does not because of cancellation explained in Appendix 2). The consequence of the lack of dependence on $x_r$ is that we can swap the order of proposal (Line~\ref{step:simple-prop}) and particle weighting (Line~\ref{step:simple-weigh}) in Algorithm \ref{algo:simple-smc}. So instead of computing the weights only for one pre-determined annealing parameter $\phi_r$, we can search over several tentative values. For each tentative value, we can score the choice using a measure of weight degeneracy applied to the putative weights. Crucially, each choice can be quickly scored without having to propose particles, which is key since proposals are typically the computational bottleneck: in a phylogenetic context, the cost of one proposal step scales linearly in the number of sites whereas the search over $\phi_r$ proposed in this section has a running time constant in the number of sites and taxa. This is because the search involves fixed values of $p(y|x_{r-1,k})$ cached from the last proposal step, which are exponentiated to different values.

Based on these observations, we select an annealing parameter $\phi$ such that we achieve a controlled increase in particle degeneracy, namely such that 
\begin{equation}\label{eq:ess-program} 
g(\phi) = \alpha g(\phi_{r-1}),
\end{equation}
where the function $g : [\phi_{r-1}, \infty) \to [0, 1]$ is defined as 
$$g(\phi) = \text{rCESS}\left(W_{r-1,\cdot}, p(y|x_{r-1,\cdot})^{\phi - \phi_{r-1}}\right),$$
and $\alpha \in (0, 1)$ is a tuning parameter, which in practice is close to $1$. By construction, $g(\phi_{r-1}) = 1$, so Equation~(\ref{eq:ess-program}) is equivalent to $g(\phi) = \alpha$.

More precisely, since we want $\phi \in [0, 1]$, the annealing parameter adaptation procedure, NextAnnealingParameterm (Algorithm~\ref{algo:Nextannealingparameter}), is designed to return $\phi_r = 1$ if $g(1) \ge \alpha$. Otherwise, because there is no closed-form solution for $\phi$ in Equation~(\ref{eq:ess-program}), we use bisection to solve this one-dimensional search problem in the interval $\phi \in (\phi_{r-1}, 1)$ (Line~\ref{line:search} of Algorithm \ref{algo:Nextannealingparameter}).

 We now argue that the search problem in Line~\ref{line:search} of Algorithm \ref{algo:Nextannealingparameter} always has a solution. Indeed, $g$ is a continuous function with, on the left end of the search interval, $g(\phi_{r-1}) = 1$, and on the right end, $g(1) < \alpha$ (otherwise the algorithm sets $\phi_{r} = 1$ in Line \ref{line:corner-case}). It follows that there must indeed be an intermediate point $\phi^*$ with $g(\phi^*) = \alpha$. Note that continuity and the identification of the left end point of the interval is possible thanks to the form of our weight update in Equation~(\ref{eq:weight-update-detail}), hence justifying the earlier informal argument about the need to have the fluctuation of the weights disappearing as $\phi_r - \phi_{r-1}$ goes to zero.

As in the previous section on dynamic resampling, NextAnnealingParameter is again based on relative conditional ESS, but this time, we are interested in the degeneracy of a single iteration, i.e. we do not trace back until the previous resampling step (since the optimization over the annealing schedule can only impact the current iteration). As a corollary, the previous iteration's particles are not always equally weighted, hence the simplification in Equation~(\ref{eq:simple-ess}) is not possible here and we use the full formula for relative conditional ESS.

\begin{algorithm}
   \caption{\bf{Procedure NextAnnealingParameter  }}
  \label{algo:Nextannealingparameter}
   {\fontsize{12pt}{12pt}\selectfont
\begin{algorithmic}[1]
  \State {\bfseries Inputs:} (a) Particle population from previous SMC iteration $(x_{r-1,\cdot}, w_{r-1,\cdot})$;
		(b) Annealing parameter $\phi_{r-1}$ of previous SMC iteration;
		(c) A degeneracy decay target $\alpha \in (0, 1)$.  

   \State {\bfseries Outputs:} automatic choice of annealing parameter $\phi_r$.

	\State  Initialize the function $g$ assessing the particle population quality associated to a putative annealing parameter $\phi$:
		\[ g(\phi) = \text{rCESS}\left(W_{r-1,\cdot}, p(y|x_{r-1,\cdot})^{\phi - \phi_{r-1}}\right)  = \frac{\left(\sum_{k=1}^K W_{r-1,k} p(y|x_{r-1,k})^{\phi - \phi_{r-1}}\right)^2}{\sum_{k=1}^K W_{r-1,k} p(y|x_{r-1,k})^{2(\phi - \phi_{r-1})}}. \] 
        
\If{$g(1) \ge \alpha$}
\State return $\phi_r = 1$. \label{line:corner-case}
  \Else
    \State return $\phi_r = \phi^* \in (\phi_{r-1}, 1)$ such that $g(\phi^*) = \alpha$ via bisection. \label{line:search}
  \EndIf
\end{algorithmic}
}
\end{algorithm}

The parameter $\alpha$ used in Algorithm~\ref{algo:Nextannealingparameter} encodes the decay in particle population quality that we are aiming for. Based on our experiments we recommend values very close to one. For this reason, we reparameterize the parameter $\alpha$ into $\alpha = 1 - 10^{-\beta}$ and recommend a default value of $\beta = 5$ as a reasonable starting point. Increasing $\beta$ improves the approximation accuracy.

\subsection{Computational Complexity}
The computational complexity of annealed SMC is linear in both the number of intermediate distributions $R$ and the number of particles $K$. Naively, the resampling step scales like $O(K^2)$, but a linear time multinomial resampling algorithm is obtained by generating order statistics via normalization of a Poisson process \citep[Section 2.1, p.214]{devroye_non-uniform_1986}. This technique is well known in the SMC literature \citep{Doucet11tutorial}. Alternatively, one can use stratified or systematic resampling \citep{Doucet11tutorial}, which provides a simple to implement linear time resampling algorithm.

The memory consumption of annealed SMC is linear in $K$ and constant in $R$.

\section{Review of other marginal likelihood estimation methods} 

For completeness, we review here some alternatives to Equation~(\ref{eq:basic-z-estimator}) for estimating marginal likelihoods, which we will compare to SMC from both a theoretical and empirical stand-point.

\subsection{Stepping Stone}

 The Stepping Stone algorithm \citep{xie2010improving} is a method for marginal likelihood estimation. It is widely used via its MrBayes implementation \citep{huelsenbeck_mrbayes:_2001}. As with SMC, the Stepping Stone method introduces a list of annealed posterior distributions connecting the posterior distribution and the prior distribution. We use a notation analogous to SMC, with $\{\pi_{d}\}_{d=0, 1, \ldots, D}$ denoting the intermediate distributions, $\pi_{d}(x)\propto  \gamma_{d}(x) = p(y|x)^{\phi_{d}}\pi(x)$, $ 0=\phi_{0} < \phi_{1}<\phi_{2}<\cdots <\phi_{D} = 1$. The marginal likelihood $Z$ can be written as  
\begin{eqnarray*}
	Z\equiv Z_D=Z_0\prod_{d=1}^D\frac{Z_d}{Z_{d-1}}. 
\end{eqnarray*}  
We can rewrite the ratio of $Z_{d}$ and $Z_{d-1}$ as 
 \begin{align}\label{eq:ratio-estimator}
	\frac{Z_d}{Z_{d-1}} = \int
	\frac{\gamma_d(x)}{\gamma_{d-1}(x)}\pi_{d-1}(x)dx.
\end{align}
The Stepping Stone method prescribes running several MCMC chains targeting $\pi_{d-1}(x)$ to obtain $N$ posterior samples $x_{d-1, 1}, x_{d-1, 2}, \ldots, x_{d-1, N}$, then 
\begin{eqnarray}\label{eq:ss-estimator}
\widehat{\frac{Z_d}{Z_{d-1}}} = \frac{1}{N}\sum_{i=1}^{N} \{p(y|x_{d-1,i})\}^{\phi_d -\phi_{d-1}}. 
\end{eqnarray} 
The estimator of the marginal likelihood admits the form
\begin{eqnarray*}
\widehat{Z_{D}} = \prod_{d =1}^{D}\frac{1}{N}\sum_{i=1}^{N} \{p(y|x_{d-1,i})\}^{\phi_d -\phi_{d-1}}. 
\end{eqnarray*} 
The number of intermediate distributions is a trade-off between computing cost and accuracy. A larger number of MCMC chains can provide a better approximation for the marginal likelihood, but the computational cost will be higher. 
To make fair comparison between the marginal likelihood estimators provided by the annealed SMC and Stepping Stone, we set $K_{SMC}R_{SMC} =N_{SS}D_{SS}$.
Another factor that will impact the Stepping Stone estimator is the choice of annealing parameter sequence $\{\phi_{d}\}_{d = 1, 2,\ldots, D}$. In this paper, we use the annealing scheme  $\phi_{d} = (d/D)^{1/a}$ recommended by \cite{xie2010improving}, where $a$ is between $0.2$ and $0.4$.

\subsection{Linked Importance Sampling}

Stepping stone uses importance sampling to approximate the ratio of marginal likelihoods for two intermediate distributions. However, the importance sampling  approximation would be poor if the two successive distributions do not have enough overlaps.  
Linked Importance Sampling  
\citep{neal2005estimating} improves the performance of importance sampling by introducing bridge distributions, e.g. ``geometric'' bridge: $\gamma_{d-1*d}(x) = \sqrt{\gamma_{d-1}(x)\gamma_{d}(x)}$. More importantly, Linked Importance Sampling  provides an unbiased marginal likelihood estimator. 
The ratio of two marginal likelihoods can be written as
\[
\frac{Z_{d}}{Z_{d-1}} = \frac{Z_{d-1*d}}{Z_{d-1}}\bigg/\frac{Z_{d-1*d}}{Z_{d}}= \bigg\{\int
\frac{\gamma_{d-1*d}(x)}{\gamma_{d-1}(x)}\pi_{d-1}(x)dx \bigg\}\bigg/\bigg\{\int
\frac{\gamma_{d-1*d}(x)}{\gamma_{d}(x)}\pi_{d}(x)dx \bigg\}.
\]
For $d = 1, \ldots, D$, to estimate the ratio $Z_{d}/Z_{d-1}$, we first run MCMC targeting $\pi_{d-1}(x)$ to obtain $N$ posterior samples $x_{d-1, 1}, x_{d-1, 2}, \ldots, x_{d-1, N}$ (when $d = 1$, we sample from the prior distribution). Then we sample the initial state of $\pi_{d}$. Two successive MCMC chains $\pi_{d-1}(x)$ and $\pi_{d}(x)$ are linked by a state $x_{d-1,\mu_{d-1}}$ where index $\mu_{d-1}$ is sampled from $\{1, 2, \ldots, N\}$ according to the following probabilities:
\[
p(\mu_{d-1}|x_{d-1,1:N}) = \frac{\gamma_{d-2*d-1}(x_{d-1,\mu_{d-1}})}{\gamma_{d-1}(x_{d-1,\mu_{d-1}})}\bigg/\sum_{i=1}^{N}\frac{\gamma_{d-2*d-1}(x_{d-1,i})}{\gamma_{d-1}(x_{d-1,i})}.
\]
In case $d = 1$, the linked state $\mu_{0}$ is uniformly sampled from the $N$ samples of $\pi_{0}(x)$. 
Finally, we run MCMC chain $\pi_{d}(x)$ starting from initial state  $x_{d-1,\mu_{d-1}}$ to obtain $N$ posterior samples $x_{d, 1}, x_{d, 2}, \ldots, x_{d, N}$. 
The ratio of two marginal likelihoods can be approximated by
\[
\widehat{\frac{Z_{d}}{Z_{d-1}}} = \widehat{\frac{Z_{d-1*d}}{Z_{d-1}}}\bigg/\widehat{\frac{Z_{d-1*d}}{Z_{d}}}
=\bigg\{\frac{1}{N}\sum_{i=1}^{N}\frac{\gamma_{d-1*d}(x_{d-1,i})}{\gamma_{d-1}(x_{d-1,i})}\bigg\}\bigg/\bigg\{\frac{1}{N}\sum_{i=1}^{N}\frac{\gamma_{d-1*d}(x_{d,i})}{\gamma_{d}(x_{d,i})}\bigg\}.
\]
In this paper, we use the ``geometric'' bridge. Hence, the estimator of ratio can be simplified to 
\begin{eqnarray*}
	\widehat{\frac{Z_d}{Z_{d-1}}} = \bigg\{\sum_{i=1}^{N} \{p(y|x_{d-1,i})\}^{\frac{\phi_d -\phi_{d-1}}{2}}\bigg\}\bigg/\bigg\{\sum_{i=1}^{N} \{p(y|x_{d,i})\}^{\frac{\phi_{d-1} -\phi_{d}}{2}}\bigg\}. 
\end{eqnarray*} 
We refer to Appendix 6 for more background on the Linked Importance Sampling algorithm.

\section{Theoretical properties}\label{sec:theory}

In this section, we review three theoretical properties of interest, \emph{consistency}, \emph{marginal likelihood estimate unbiasedness}, and \emph{asymptotic normality}, with an emphasis on their respective practical importance.

\subsection{Properties of Annealed Sequential Monte Carlo}

In the context of SMC algorithms, the first property, \emph{consistency} means that as the number of particles is increased, the approximation of posterior expectations can become arbitrarily close to the true posterior expectation. This makes the approximation in Equation~(\ref{eq:expectation-approx}) more precise:
\begin{equation}\label{eq:consistency} 
\sum_{k=1}^K W_{r,k} f(x_{r,k}) \to \int \pi_r(x) f(x) \ud x \;\text{as $K\to\infty$,}
\end{equation}
provided $f$ satisfies regularity conditions, for example $f$ is bounded, and where convergence of the random variables holds for a set of random seeds having probability one. See for example \cite{LiangliangWang2015}. 

Consistency can be viewed as the ``bare minimum'' expected from modern SMC algorithms. A more informative class of results consists in central limit theorem equivalents of Equation~(\ref{eq:consistency}). These results can be used to assess the total variance of Monte Carlo estimators (whereas measures such as effective sample size described previously are local in nature), see \cite{chan_general_2013}. However, since numerically stable versions of these methods are still at their infancy \citep{olsson_numerically_2017}, we will focus the remaining on the third property, unbiasedness.   

We say an estimator $\hat Z$ for a constant $Z$ is unbiased if $\E[\hat Z] = Z$. Here the expectation is defined with respect to the randomness of the approximation algorithm. This contrasts with the classical statistical definition of unbiasedness in which the randomness comes from the data generation process.

For SMC algorithms, unbiasedness holds in a more restrictive sense compared to consistency. In general:
\begin{equation}\label{eq:bias} 
\E \left[ \sum_{k=1}^K W_{r,k} f(x_{r,k}) \right] \neq \int \pi_r(x) f(x) \ud x,
\end{equation}
in other words, repeatedly running SMC with a fixed number of particles but different random seeds and averaging the results does not provide arbitrarily precise approximations (the same negative result holds with MCMC). 
However, if we restrict our attention to marginal likelihood estimates, remarkably the unbiasedness property does hold \citep{del2006sequential}, i.e. for any finite $K$, $\hat Z_K$ as defined in Equation~(\ref{eq:basic-z-estimator}) is such that:
\begin{equation}\label{eq:unbiasedness} 
\E \left[ \hat Z_K \right] = Z = \int \gamma_r(x) \ud x.
\end{equation}
More details on the unbiasedness of the  marginal likelihood SMC estimator and other theoretical properties of annealed SMC can be found in Appendix $2$, subsection \emph{Unbiasedness, Consistency and Central Limit Theorem}.

While the notion of unbiasedness has been central to frequentist statistics since its inception, only in the past decade has it started to emerge as a property of central importance in the context of (computational) Bayesian statistics. Traditionally, the main theoretical properties analyzed for a given Monte Carlo method  $\hat Z$ estimating $Z$ was consistency.

 With the emergence of pseudo-marginal methods,  the bias of Monte Carlo methods is now under closer scrutiny. Pseudo-marginal methods are MCMC methods which replace probability factors in the Metropolis-Hastings ratio by positive unbiased estimators of these probabilities. For example, \cite{Andrieu2010} provide examples where global parameters of state-space models are sampled using an MCMC algorithm where the probability of the data given the global parameters and marginally over the latent states is estimated using an SMC algorithm. 
We refer the reader to \cite{andrieu_pseudo-marginal_2009} for  more examples where unbiasedness is used to compose MCMC algorithms in order to attack inference in complex models. In the context of phylogenetic inference, this is useful for Bayesian analysis of intractable evolutionary models, see for example \cite{Hajiaghayi2014Efficient}. 

Another area where unbiasedness can play a role is for checking correctness of Monte Carlo procedures. In contrast to correctness checks based on consistency such as \cite{geweke_getting_2004}, which are asymptotic in nature and hence necessarily have false positive rates (i.e. cases where the test indicates the presence of a bug when in fact the code is correct), checks based on unbiasedness can achieve a false positive rate of zero, using the strategy described in the next section.

\subsection{Using Unbiasedness to Test Implementation Correctness}\label{sec:correctness-test}

Typically, the algorithm shown in Algorithm \ref{algo:simple-smc} is implemented in a model-agnostic fashion. Hence it is reasonable to assume that we can construct test cases on discrete state spaces. For example, one can use phylogenetic trees with fixed branch lengths, or even simpler models such as hidden Markov models (HMMs). Furthermore, we conjecture that many software defects can be detected in relatively small examples, where exhaustive enumeration is possible, and hence $Z$ can be computed exactly. We can determine sufficient complexity of the examples to use via code coverage tools \citep{miller_systematic_1963}. 

We would like to test if equality of Equation~(\ref{eq:unbiasedness}) holds for a given implementation. The right hand side can be computed easily since we assume the example considered is small. To compute analytically the expectation on the left-hand side, we use a method borrowing ideas from probabilistic programming \citep{wingate_nonstandard_2011}, and use an algorithm, called ExhaustiveRandom that automatically visits all possible execution traces $\tau_i$ of a given randomized algorithm.
The execution trace of a randomized algorithm refers to a  realization of all random choices in the algorithm (in the context of SMC, both the resampling steps and the proposal steps). ExhaustiveRandom enumerates all the execution traces while also computing the respective probability $p_i$ of each trace. This is done by performing a depth first traversal of the decision tree corresponding to the randomized algorithm being tested. The number of execution traces grows exponentially fast but this is still a useful tool as very small examples are generally sufficient to reach code coverage. 

For each execution trace $\tau_i$, we can also obtain the normalization estimate $\hat z_i$ corresponding to that trace, and hence get the value of the left-hand side of Equation~(\ref{eq:unbiasedness}) as $\sum_i p_i \hat z_i$. We used this check via an open source implementation of ExhaustiveRandom ({\small   \url{https://github.com/alexandrebouchard/bayonet/blob/1b9772e91cf2fb14a91f2e5e282fcf4ded61ee22/src/main/java/bayonet/distributions/ExhaustiveDebugRandom.java}}) to ensure that our software satisfies the unbiasedness property. See the numerical simulation section for details.

\subsection{Properties of the Stepping Stone Method}

For the stepping stone method, the expected value of Equation~(\ref{eq:ss-estimator}) depends on the nature of the samples $x_{d-1,1}, x_{d-1, 2}, \dots, x_{d-1,N}$. If they are independent, the procedure is unbiased. However, if the samples are obtained from a Markov chain, there are no guarantees that the procedure is  unbiased unless the MCMC chain is initialized at the exact stationary distribution.  In practice, this is not possible: \cite{xie2010improving} use a burned-in MCMC chain, which implies that  the chain is \emph{asymptotically unbiased}, however for any finite number of iterations, a bias remains. Unfortunately, the two main motivations for unbiasedness (pseudo-marginal methods and the correctness checks described earlier) both require unbiasedness to hold for any finite number of Monte Carlo samples; asymptotic unbiasedness is not sufficient.

We show in the numerical simulation section an explicit counterexample where we compute the non-zero bias of the stepping stone method. This motivates the need for implementable unbiased methods, such as the annealed Sequential Monte Carlo method described in this work.

\subsection{Comparison of Unbiased Marginal Likelihood Estimators}

In Bayesian phylogenetics, the marginal likelihood estimate is generally a very small number. Instead of computing $\hat{Z}$ directly, we compute the logarithm of  the marginal likelihood estimate, $\log(\hat{Z})$. For SMC and Linked Importance Sampling, although $\hat{Z}$ is an unbiased estimator, taking the logarithm of $\hat{Z}$ introduces bias. Jensen's inequality shows that $\log(\hat{Z})$ is a biased estimator of $\log(Z)$, and is generally underestimated,
\[
\E[\log(\hat{Z})] \leq \log(\E(\hat{Z})) = \log Z.
\]

This provides a tool to compare the performance of an unbiased normalization constant estimation method $m_1$ to another one $m_2$. Suppose we run each method $M$ times with different seeds and a fixed computational budget. Let $L_i = \sum_{j=1}^M \log \hat Z_{i,j} / M$ denote the average estimate of the log marginal likelihood for the $i$-th method, where $\hat Z_{i,j}$ is the estimate  with the $j$-th random seed using the $i$-th method. If $m_2$ is also unbiased then for $M$ large enough,  both $m_1$ and $m_2$ underestimate $\log \E[Z]$, and the largest $L_{i}$ is closest to $\log \E[Z]$, which determines the best performing method. If $m_2$ is not unbiased, then if $L_1 > L_2$ and $M$ is large enough, we can conclude that $m_1$ is superior (but we cannot confidently order the methods if $L_2 > L_1$). 

However the Monte Carlo counterparts of the orderings should be considered with a pinch of salt since the number of replicates $M$ needed may be intractable in some cases.

\section{Simulation Studies}

\subsection{Simulation Setup and Tree Distance}

In order to simulate datasets, we first generated a set of
random unrooted trees, including topology and branch lengths, as the reference trees.  The tree topology was sampled from a uniform distribution. Each branch length was generated from an exponential distribution with rate 10.0. 

Then, for each reference tree,  we simulated DNA sequences  using the {K2P}
model with parameter $\kappa=2.0$ \citep{kimura1980simple}. While the main focus of this work is on marginal likelihood estimation, we also performed some benchmarking on the quality of the inferred trees. 
To do so, we used the \emph{majority-rule consensus tree} \citep{felsenstein1981} to summarize the weighted phylogenetic tree samples obtained from annealed SMC. 
We measured the distance between each estimated consensus
tree to its associated reference tree using three types of distance metrics:  the Robinson-Foulds ({RF}) metric based on sums of differences in branch lengths  \citep{RobinsonFoulds1979}, the Kuhner-Felsenstein ({KF}) metric
\citep{Kuhner01051994}, and the partition metric ({PM}), also known as symmetric difference or topology only RF metric \citep{Robinson1981}. 

\subsection{Hidden Markov Models}

\begin{figure}[htbp]
	\centering
	\includegraphics[width=1\textwidth]{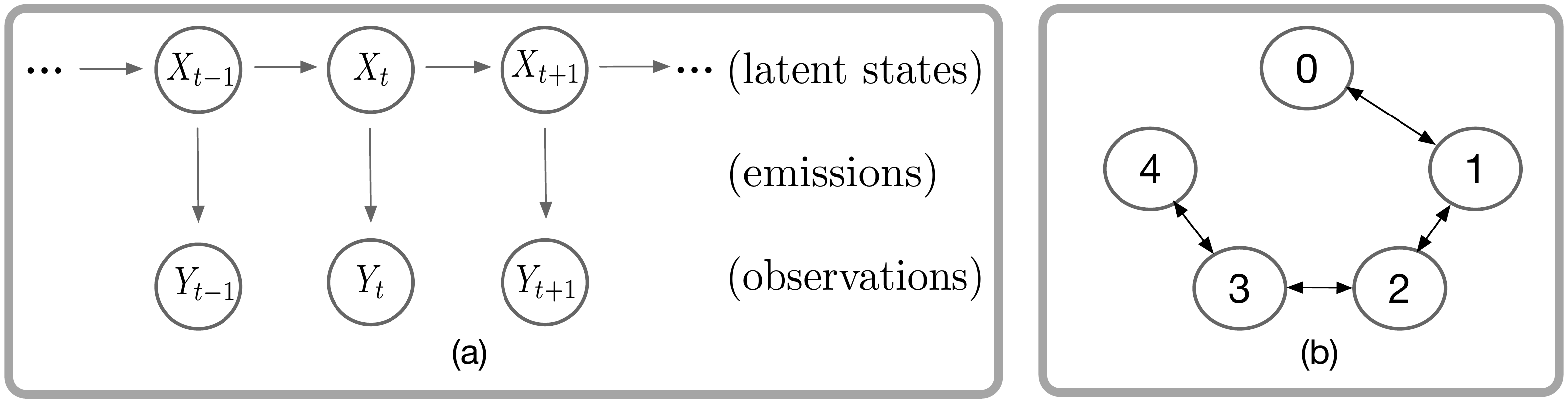}
	\caption{Graphical representation of (a) a hidden Markov model; (b) transitions between hidden states. }
	\label{fig:hmmgraph}
\end{figure}

As discussed when we introduced the unbiasedness correctness test, it is useful to perform some preliminary experiments on finite state models. We used a hidden Markov model (HMM) with a finite latent state (see the graphical representations of a hidden Markov model and hidden state transitions in Figure \ref{fig:hmmgraph}). The variables $X_t$ shown in the figure are unobserved and take on discrete values with a distribution depending on the previous variable $X_{t-1}$. For each unobserved variable, we define an observed variable $Y_t$, also discrete, with a conditional distribution depending on $X_t$. The latent state space in our experiment was set to $\{0, 1, 2, 3, 4\}$ and latent transitions were set uniformly across neighbour integers. The emissions we used take two possible values with conditional probabilities given by $(0.2,0.8), (0.1, 0.9), (0.01,0.99), (0.2, 0.8)$ and $(0.3,0.7)$. The proposals were based on the Gibbs sampler on a single variable. The posterior distribution of interest is over the latent variables $X_{1}, X_2, \dots$ given the observations $Y_1, Y_2, \dots$. Of course, such a model would not normally be approached using approximate inference methods. Moreover, notice that this is a non-standard way of using SMC for a sequential model where we do not make use of the sequential structure of the model.

We first performed unbiasedness correctness tests on a chain of length two based on three equally spaced annealing parameters $(0, 1/2, 1)$, and observations $(0, 1)$. We first computed the true marginal likelihood, $0.345$. Using the method described in the Theoretical Properties Section, we computed the exact value of $\E[\hat Z]$ by exhaustive enumeration of all execution traces for SMC and the Stepping Stone method. For SMC with two particles, the ExhaustiveRandom algorithm enumerated $1,992,084$ traces resulting in an expectation of $0.34499999999999525$. For Stepping Stone with two MCMC iterations per annealing parameter, the ExhaustiveRandom algorithm enumerated $1,156,288$ traces resulting in an expectation of $0.33299145257312235$. This supports that SMC is unbiased and provides an explicit counterexample of the bias of the stepping stone method.

\begin{figure}[htbp]
	\centering
	\includegraphics[width=1\textwidth]{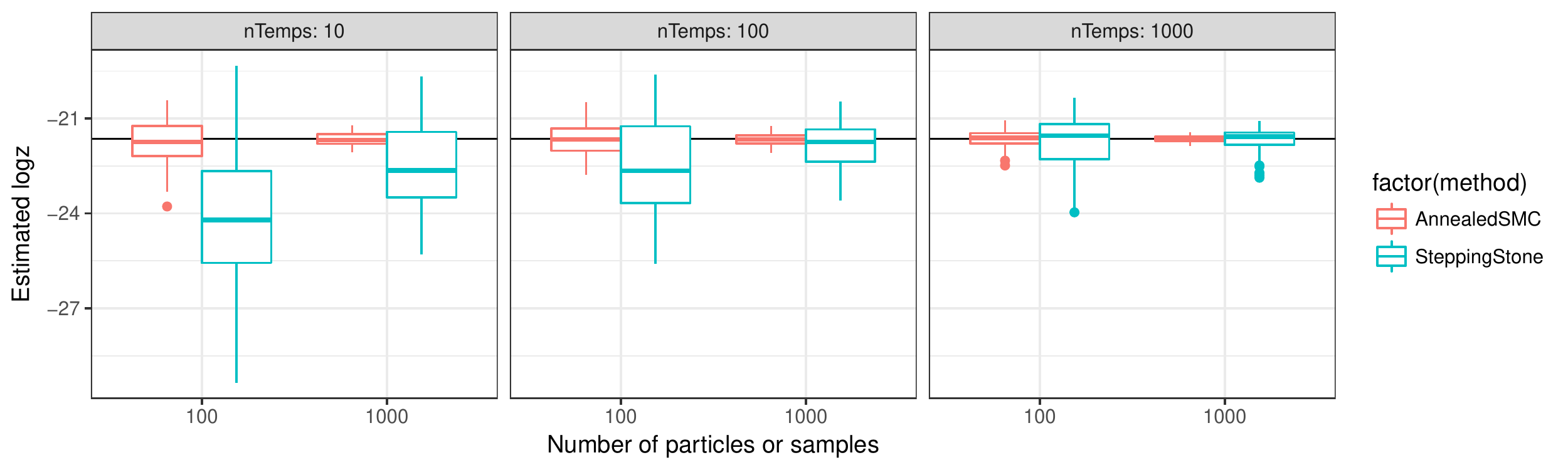}
	\caption{Estimates of the log marginal likelihood based on 100 independent random seeds for each configuration.  The axis label ``nTemps'' refers to the number of equally spaced intermediate distributions (``inverse temperatures''). The black line shows the true value computed using the forward-backward algorithm. \label{fig:hmm}}
\end{figure}

Second, we ran experiments on larger versions of the same model, a chain of length 32, as well as with more annealing steps and particles per step. In this regime it is no longer possible to enumerate all the execution traces so we averaged over 100 realizations of each algorithm instead. The true marginal likelihood can still be computed using a forward-backward algorithm. We show the results in Figure~\ref{fig:hmm}. 

\subsection{Comparison of Marginal Likelihood Estimates}

In this section, we benchmark the marginal likelihood estimates provided by adaptive annealed SMC (ASMC), debiased adaptive annealed SMC (DASMC), deterministic annealed SMC (DSMC), Linked Importance Sampling (LIS) and Stepping Stone (SS). In DASMC, the annealing scheme was determined before running annealed SMC using the same annealing parameters obtained from the ASMC.
In DSMC, we used the annealing scheme $\phi_{r} = (r/R)^{3}$ with a predetermined $R$. 

In the first experiment, we focus on evaluating the marginal likelihood estimates using ASMC, DASMC, LIS and SS with the same computing budget.  
We simulated unrooted trees of varying sizes (numbers of taxa): $5$, $10$, $15$, $20$, and $25$. 
For each tree, we generated one data set of DNA sequences. Sequence length was set to $100$. The execution of each algorithm and setting was repeated $100$ times with different random seeds. We used $\beta = 5$ for adaptive annealed SMC, and the number of particles was set to $1000$. 
In stepping stone and linked importance sampling, we set the total number of heated chains $D$ to 50, and the annealing scheme was set to $\phi_{d} = (d/D)^{3}$, where $d = 1, 2, \ldots, D$. We enforced $K_{SMC}R_{SMC} = N_{SS}D_{SS} = N_{LIS}D_{LIS}$ in order to make the comparisons fair. Information about $R_{SMC}$ are shown in  Table \ref{tab: R}.

\begin{table}[htbp]
   \caption{Number of annealing parameters ($R_{SMC}$) for ASMC with $K=1000$ and $\beta = 5$.}   \label{tab: R}
   \centering
   \begin{tabular}{l|llllll}           
 \#taxa & $5$ & $10$ & $15$ & $20$ & $25$ \\ \hline
$R_{SMC}$  &   $1932$ & $3741$ & $5142$ & $6047$& $7219$\\
\end{tabular}
\end{table}

\begin{figure}[htbp]
  \centering
\includegraphics[width=1\textwidth]{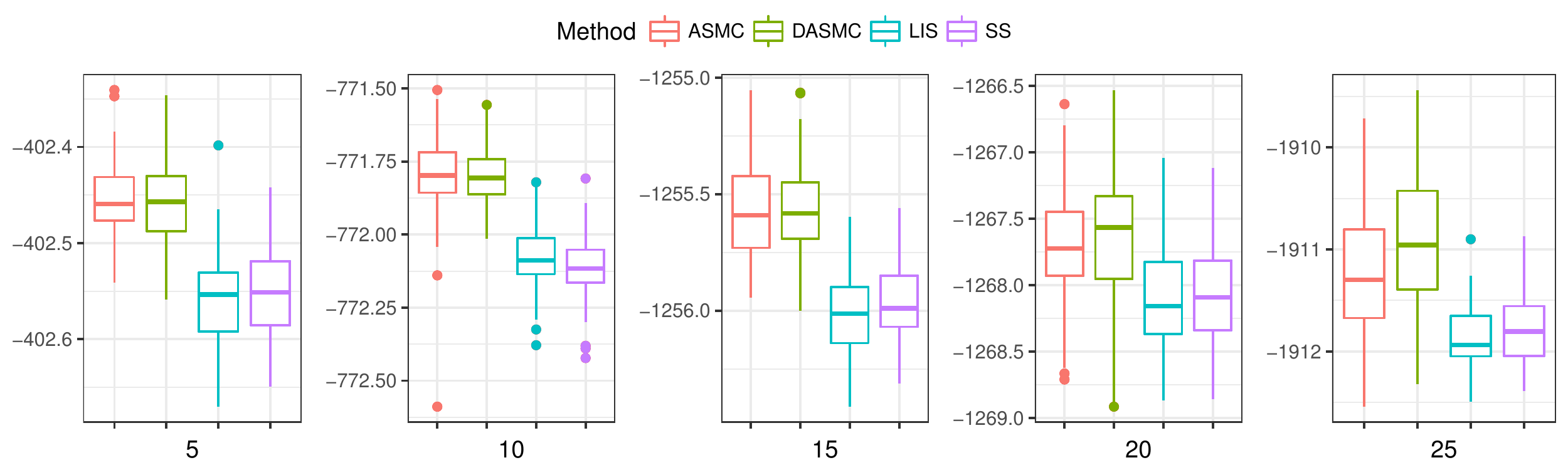}
\caption{Marginal likelihood (in log scale) estimates for different numbers of taxa with a fixed computational budget.}
 \label{fig:logZSMCMCMC}
 \end{figure}

Figure \ref{fig:logZSMCMCMC} shows the comparison of the performance of the four algorithms in terms of the marginal likelihood in log scale as the number of taxa increases.  As described in the theoretical analysis section, for the unbiased estimators, we have asymptotically that the log of the marginal likelihood should underestimate the marginal likelihood by Jensen's inequality.  The results support that ASMC and DASMC can achieve more accurate marginalized likelihood estimates compared to SS and LIS with the same computational cost. The performances of the two SMC algorithms are quite similar, while the marginal likelihood estimates provided by LIS and SS are close to each other.
 In Appendix 9, we describe an experiment comparing ASMC, DASMC, LIS and SS with a very large value of $K$. The mean of log marginal likelihood for the four methods are close (reduction of the gap is expected, since all methods are consistent), while ASMC and DASMC still exhibit smaller variance across seeds compared to LIS and SS. 

Another experiment was conducted to measure the variability of the marginal likelihood estimates from each algorithm,  
by comparing the coefficients of variation (CV) for different numbers of taxa with the same setting.  The coefficient of variation is defined as $CV = sd(\hat{Z})/\E(\hat{Z})$.  
 We simulated $70$ trees, increasing the number of taxa (from 10, 15, 20, 25, 30, 35, 40; 10 trees of each size), and created $10$ data sets for each tree.  For each data set, we repeated each algorithm $10$ times with different random seeds. The upper bound of CV equals $\sqrt{n-1}$, where $n$ represents the number of repeats with different random seeds in experiments. We refer to Appendix 7 for the derivation of the upper bound of CV. In our setting, this upper bound is $\sqrt{10-1} = 3$. In ASMC, the computational cost was fixed at $K = 1000$ and $\beta = 5$. In DSMC, we used the same number of particles, and the annealing scheme was set to  $\phi_{r} = (r/R)^{3}$, where the total number of annealing parameters $R$ was fixed to be the one obtained from running ASMC with $K = 1000$ and $\beta = 5$ for a tree with $10$ taxa. 

\begin{figure}[htbp]
  \centering
\includegraphics[width=0.45\textwidth]{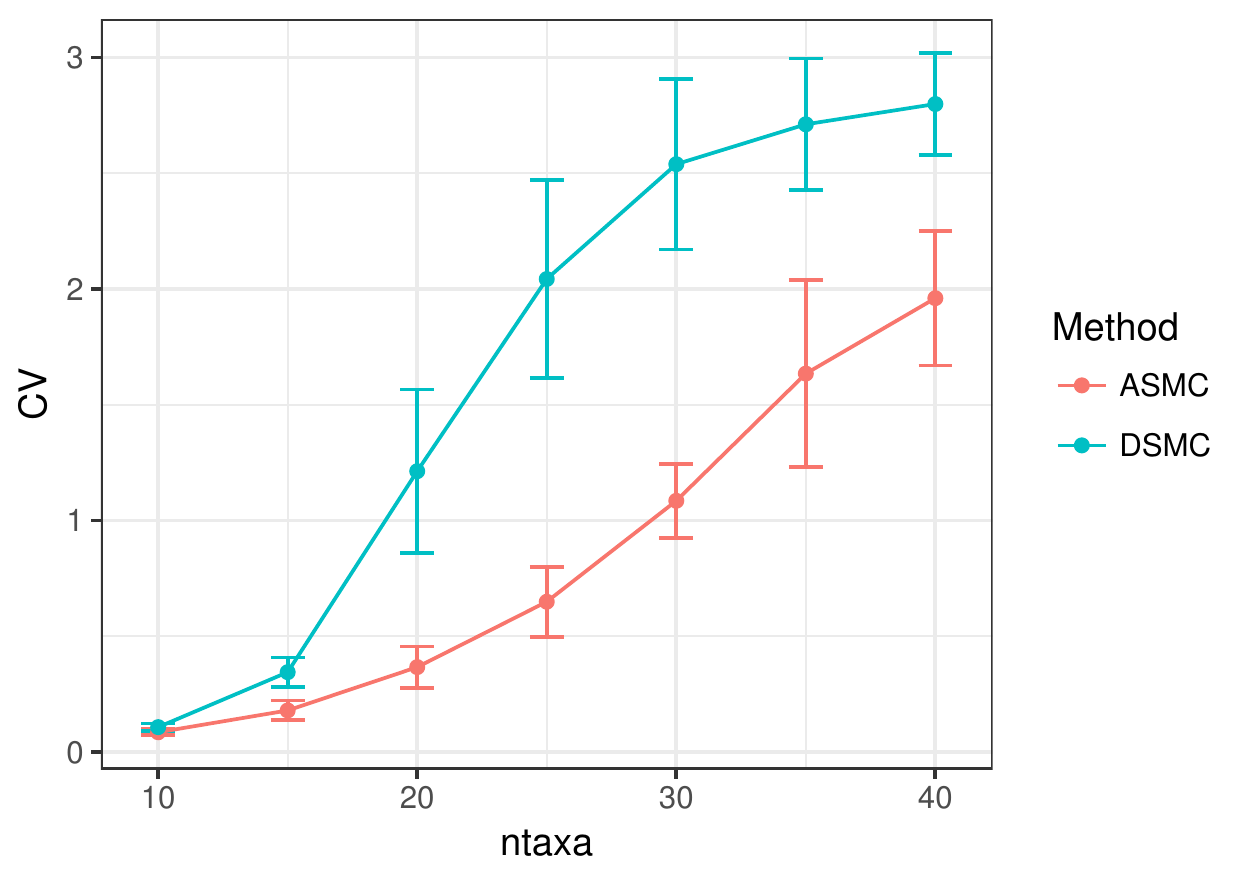}
\caption{Coefficient of variation (CV) for the marginal likelihood estimates versus the number of taxa for ASMC and DSMC with a fixed number of particles.}
 \label{fig:CV_SMC}
 \end{figure}

Figure \ref{fig:CV_SMC} displays the CV for ASMC and DSMC as a function of the number of taxa. The error bars in the figure represent $95\%$ confidence intervals. The CV of DSMC increases faster than ASMC as the number of taxa gets larger than $15$. It gradually converges to the upper bound of CV as the number of taxa reaches $35$. The CV of ASMC increases more slowly as the number of taxa increases. 

\subsection{Comparison of Model Selection by Annealed SMC versus Stepping Stone}
In this section, we compare the performance of ASMC and Stepping Stone on a Bayesian model selection task.  
We simulated 20 unrooted tree of 10 taxa using a uniform distribution for the tree topology and branch lengths generated from an exponential distribution with rate 10. A total of sixty data sets of DNA sequences of length 500 were generated using each of the simulated tree and the following three evolutionary models: JC69, K2P, and GTR+$\Gamma$. The parameter $\kappa$ in the K2P model was set to 2.0.  In the GTR+$\Gamma$ model, a symmetric Dirichlet distribution with parameters $(10,10,10,10)$ was used to generate the base frequencies, and a symmetric Dirichlet with parameters $(10,10,10,10,10,10)$ was used to generate the GTR relative rates in the rate matrix. The discrete gamma distribution with 4 categories was used to convey among-site rate heterogeneity, with the gamma shape parameter drawn from a Gamma distribution with parameters $(2,3)$.   

For each data set, marginal likelihoods were estimated by ASMC and SS using three evolutionary models, JC69, K2P, and GTR, respectively.  In ASMC, we used $K=1000$ and $\beta=4$.  The total number of iterations in SS was set to the product of the number of particles and number of iterations in ASMC.   
Table \ref{tab: modelSelection} shows the Bayesian model selection results.  
Both ASMC and SS choose the correct model for all of the 20 data sets generated from the JC69 and K2P model, respectively. For the data generated from GTR+$\Gamma$, SMC chooses the closest model, GTR, 18 times out of 20, while SS only chooses GTR 15 times out of 20. 

\begin{table}[htbp]
   \caption{Comparison of model selection by ASMC and SS based on the Bayes factor.}   \label{tab: modelSelection}
   \centering
   \begin{tabular}{ll|llllll}           
   && \multicolumn{3}{c}{Data generated from} \\
Method & Model & JC69 & K2P & GTR+$\Gamma$  \\ \hline
ASMC&JC69   &   20 & 0 & 1 \\
&K2P   & 0  & 20 & 1 \\
&GTR   & 0  & 0 &  18\\ \hline
SS&JC69   &   20 & 0 & 4 \\
&K2P   &  0 & 20 & 1 \\
&GTR   &  0 & 0 &  15\\
\end{tabular}
\end{table}

\subsection{Comparison of Tree Distance Metrics}

In this section, we compare the quality of reconstructed phylogenies using synthetic data.  
We simulated one unrooted tree, the reference tree, with $50$ taxa and then generated one data set of DNA sequences of length $2000$ from this tree. 
The ASMC was run with $\beta = 6$ and $K = 100$. 
The MCMC algorithm was initialized with a random tree 
from the prior distribution.  To make a fair comparison, we set the number of MCMC iterations to be no less than $K_{SMC}R_{SMC}$. We discarded 20\% of the MCMC chain as ``burn-in''. Table \ref{tab:Tree502000} 
summarizes the iteration numbers, the log likelihood of the consensus tree and tree distance metrics from running ASMC and MCMC.    Although the computational cost of MCMC is set to about twice as high as ASMC, 
the log-likelihood of the consensus tree from ASMC is much higher than that from MCMC. In addition, ASMC achieves much lower RF and KF distances to the reference tree. 
Further, to confirm that both 
ASMC and MCMC can converge to the same posterior distribution, MCMC was rerun with a better starting value, namely the consensus tree obtained after running ASMC. 
This run of MCMC is denoted as MCMC2 in Table \ref{tab:Tree502000}. The computational cost of MCMC2 is set the same as the ASMC algorithm. This time MCMC achieved similar consensus tree log-likelihood and tree distance metrics compared to ASMC, which supports that MCMC is indeed ``trapped'' in a sub-space.

\begin{table}[htbp]
   \caption{Comparison of tree distance metrics using ASMC and MCMC.}
   \label{tab:Tree502000}
   \centering
   \begin{tabular}{lllll}          
Method	 &	$R$  &	$K$	&	Metric  &	Value	 \\ \hline
ASMC   	& 	54876	&  100	&  ConsensusLogLL	 & -72787.99	 \\
&	54876	&  100	&	BestSampledLogLL   &	-72826.17	\\
& 54876	&  100	&	PartitionMetric	         &    0	                \\
& 54876	&  100  	&	RobinsonFouldsMetric &	0.70623	\\
& 54876	&  100	&	KuhnerFelsenstein	 & 0.00990	 \\
 \hline
MCMC		&		1.0E+07	& &	ConsensusLogLL &	-72833.82	\\
		&	1.0E+07		& &	PartitionMetric	&	0	 \\
		&	1.0E+07		&&	RobinsonFouldsMetric &	0.92031  \\
		&	1.0E+07		&&	KuhnerFelsenstein &	0.03138\\
 \hline
MCMC2		&		5.49E+06	& &	ConsensusLogLL &	-72784.86	\\
		&	5.49E+06		& &	PartitionMetric	&	0	 \\
		&	5.49E+06		&&	RobinsonFouldsMetric &	0.73644\\
		&	5.49E+06		&&	KuhnerFelsenstein &	0.01066
\end{tabular}
\end{table}

\subsection{Influence of Number of Threads, $\beta$, and $K$}

The runtime of the ASMC is dependent on the 
the number of threads, as well as on the tuning parameters $\beta$ and $K$. In this section, we focus on investigating the effects of these factors on ASMC.  We simulated an unrooted tree with $30$ taxa at the leaves, and then generated DNA sequences of length $1500$. 

\begin{figure}[htbp]
 \centering
\includegraphics[width=0.4\textwidth]{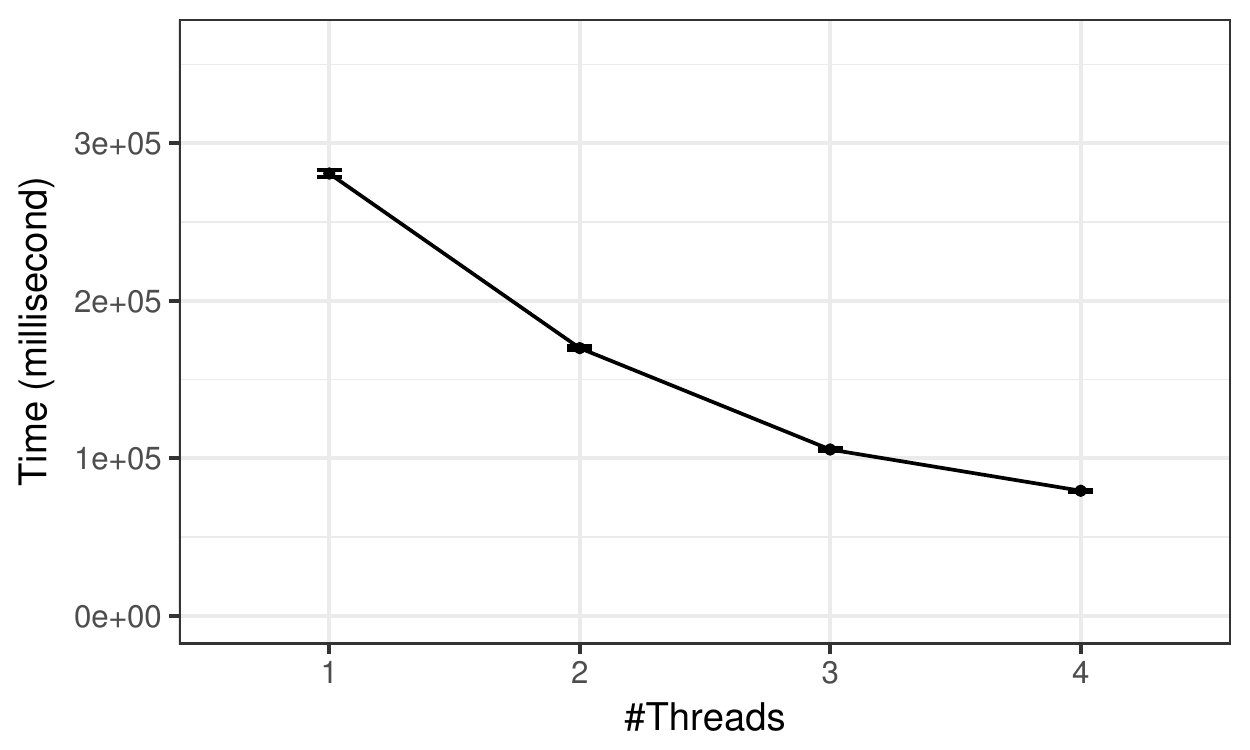}
\caption{Computing time of ASMC using multiple threads.}
\label{fig:threads}
\end{figure}

Next, Figure \ref{fig:threads} displays the computing time versus number of threads for an implementation of ASMC where the proposal step is parallelized. The error bars represent the $95\%$ confidence intervals based on 100 runs. We used $K = 1000$ and $\beta = 2$ for each number of threads. The results indicate that by increasing the number of cores, the speed of the ASMC algorithm can be increased notably. 

In Table  \ref{tab: betaandK},  we compare the performance of ASMC algorithm as a function of $K$, with $\beta$ fixed at $5$. We chose four different particle values $K = 100, 300, 1000, 3000$. The marginal likelihood estimates improve as $K$ increases. 
	
We also compared the performance of ASMC algorithm as a function of $\beta$, with $K = 1000$. 
We selected four distinct $\beta$ values, $\beta = 3, 4, 5, 5.3$.
As expected, the marginal likelihood estimates improve when $\beta$ increases. 
The likelihood of the consensus trees and tree distance metrics provided by these two experiments are displayed in Appendix 8. In practice, a value of $\beta$  close to 5 is recommended as the default value. 

\begin{table}[htbp]
   \caption{Comparison of adaptive SMC algorithm with different numbers of particles and $\beta$.}
   \label{tab: betaandK}
   \centering
   \begin{tabular}{l|l|l|l}          
$K$ $(\beta = 5)$	 &	\hspace{20mm}  $\log(Z)$ &	$\beta$ $(K = 1000)$	&	\hspace{20mm}  $\log(Z)$  	 \\ \hline
$100$   	& 	-28,288.5 (-28,283.9, -28,293.7)	&  3	&  -28,524.8 (-28,466.2, -28,641.0)	 	 \\
$300$&-28,283.5 (-28,281.1, -28,287.9)	&  4	&	-28,312.2 (-28,304.5, -28,328.8)   	\\
$1000$& -28,280.5 (-28,278.3, -28,283.5)	&  5	&	-28,280.5 (-28,278.3, -28,283.5)       	                \\
$3000$& -28,279.3 (-28,278.1, -28,280.4)	&  5.3  	&	-28,279.5 (-28,278.7, -28,280.5 ) 	\\
\end{tabular}
\end{table}

\subsection{Trade-off between $R$ and $K$}
\begin{figure}
    \centering
        \includegraphics[width=0.8\textwidth]{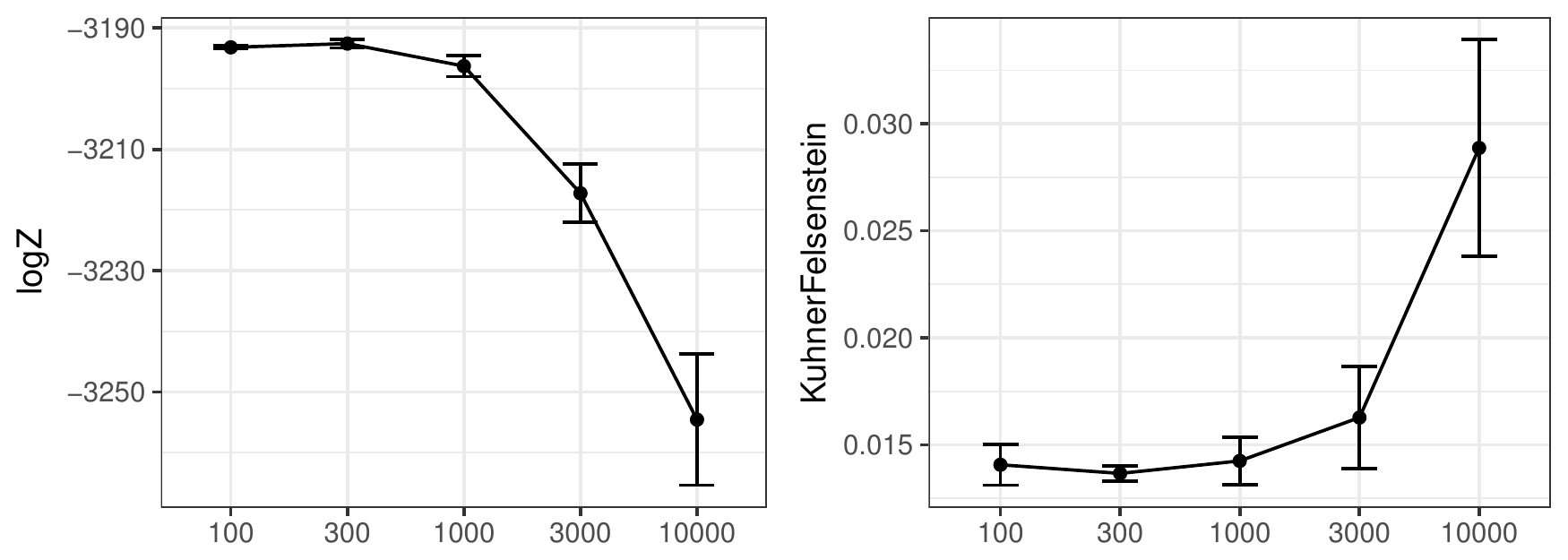}
        \caption{Performance of deterministic SMC algorithm on a fixed computational budget $(K\cdot R=10^6)$.   We select $5$ values of $K$,  $100, 300, 1000, 3000, 10000$, from left to right on X-axis. }
        \label{fig:SMCsquare}
\end{figure}

We conducted an experiment to investigate, for a given amount of computation, the relative importance of $R$ and $K$ in improving the quality of the posterior distribution inferred by annealed SMC. We used DSMC with a cubic annealing scheme. We selected values for the tuning parameter $K$ $(100, 300, 1000, 3000, 10000)$, and for each value of $K$, a corresponding value for $R$ such that the total computation cost $K\cdot R$ is fixed at $10^{6}$. We simulated one unrooted tree of $15$ taxa, and  generated one data set of DNA sequences. Sequence length was set to $300$. Figure  \ref{fig:SMCsquare} displays the marginal likelihood estimates, and KF metric provided by DSMC with different $K$ values when the total computational budget $(K\cdot R)$ is fixed.    
This results indicate that for a given amount of computation, a relatively small $K$ and a large $R$ is optimal. However, the value of $K$ cannot be too small, as an extremely small $K$ necessarily leads to a large Monte Carlo variance. 

\subsection{Analysis of Subsampling SMC}

Subsampling SMC, detailed in Appendix 1, can be used to speed up the SMC algorithms at the cost of decreasing the accuracy of estimation. The idea is to divide the data, the sites of biological sequences in our case, into batches, and only use a subset of the data in the intermediate distributions.  In this section, we evaluate the impact of the batch size (the number of sites of biological sequences in each batch), denoted $b_s$, on the speed of the algorithm and the posterior approximation. 

\begin{figure}
   \centering
    \includegraphics[width=0.5\textwidth]{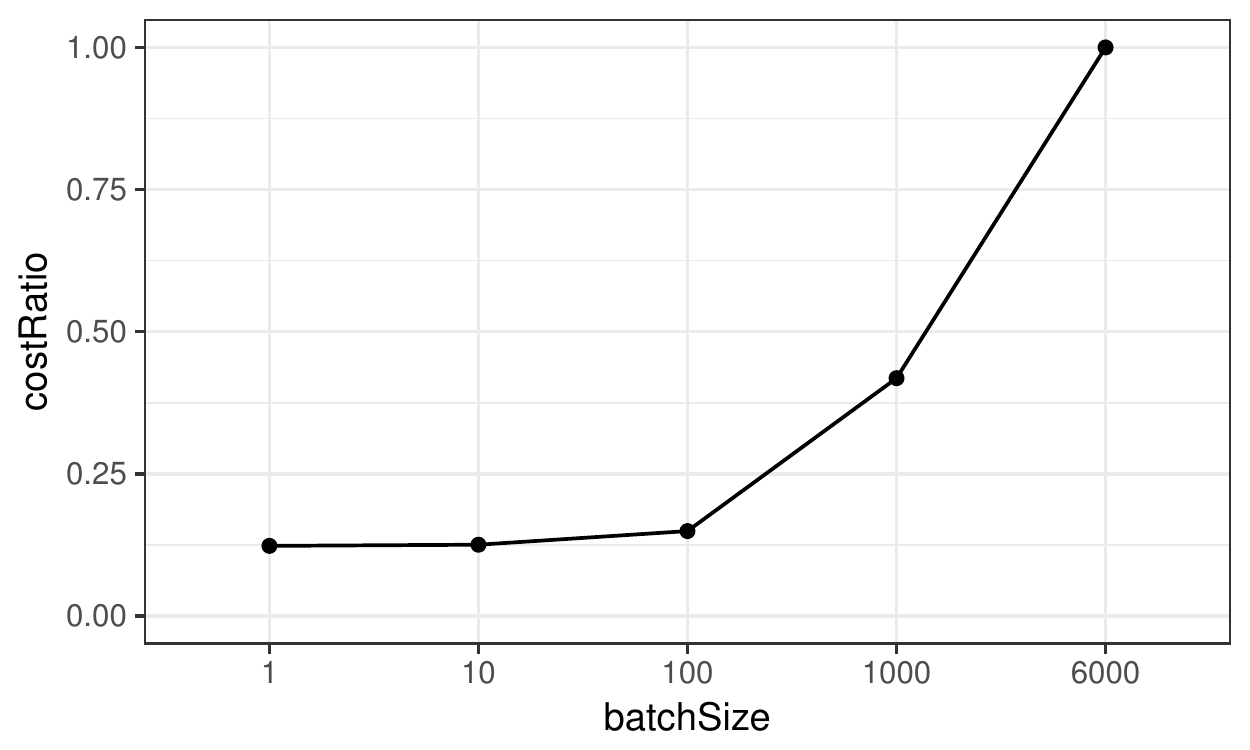}
     \caption{Ratio of cost  (subsampling/annealing) versus the batch size.}
     \label{fig: costRatio}
\end{figure}

In a first experiment, we analyzed the relative computational cost of subsampling SMC with respect to annealed SMC for different batch sizes.  
We simulated an unrooted tree with $10$ taxa, and then  generated DNA sequences of length $6000$.
The annealing parameter sequence $\phi_{r}$, $r = 0, 1, \ldots, R$,  was chosen by 
 running adaptive ASMC using $\beta = 4$ and $K = 100$. The computational cost in this subsection is measured by the total number of  sites involved in computing the unnormalized posterior and the weight update function. For example, in this simulation study, the total number of annealing parameters in adaptive ASMC is $2318$, and the number of  sites involved in each SMC iteration is $6000$. Using the fact that the likelihood for particles evaluated at iteration $r-1$ can be used to evaluate the weight update function at iteration $r$,  the total cost for ASMC is $2318\cdot 6000 = 1.39\times 10^{7}$. Figure \ref{fig: costRatio} displays the ratio of computational cost (subsampling/annealing) versus the  batch size.  The cost ratio increases slowly when we increase the batch size from $1$ to $100$. 

\begin{figure}
   \centering
    \includegraphics[width=1\textwidth]{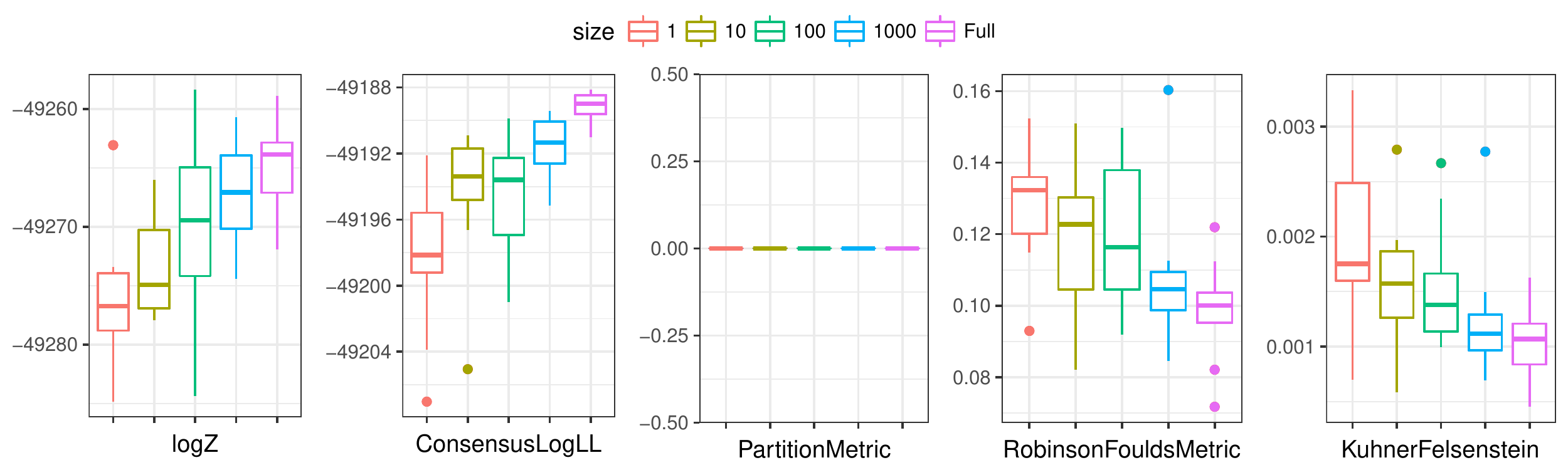}
     \caption{Comparison of subsampling SMC algorithms with different size of batch sites.}
     \label{fig: subsamplingComparison}
\end{figure}

We investigated the performance of subsampling SMC with different  batch sizes, $b_{s} = 1, 10, 100, 1000, 6000$ in terms of phylogenetic tree inference. We used $K = 100$ and  ran the  subsampling SMC algorithm $10$ times for each value of $b_{s}$. The schedule $\phi_{r}$ used to compute the annealing parameter $\psi(s, \phi_{r})$ in subsampling SMC was obtained by running adaptive annealed SMC once using $\beta = 4$ and $K = 100$.
Figure \ref{fig: subsamplingComparison} displays the performance of the subsampling algorithm with different $b_{s}$. As expected, there is a trade-off between the computational cost and accuracy of most metrics: for all metrics except the partition metric, subsampling produces lower quality approximations at a lower cost. However, if the user only require a reconstruction of the tree topology, the partition metric results provide an example where subsampling is advantageous. 

\subsection{Comparison of ASMC and Combinatorial SMC (CSMC)}

We compared the performance of the annealed SMC and the combinatorial SMC algorithm (CSMC) \citep{LiangliangWang2015} for three different kinds of trees: clock, relaxed clock, and nonclock. The clock trees were simulated by assuming that the waiting time between two coalescent events is exponentially distributed with rate 10. The relaxed clock trees were obtained by perturbing the branch length of clock trees. More specifically, we modified each branch of length $l$ by adding to it a noise randomly sampled from Unif(-0.3$l$, 0.3$l$). The nonclock trees were simulated with uniformly distributed tree topologies and exponentially distributed branch lengths with rate 10. 

For each type of phylogenetic tree, we simulated 10 trees with 10 leaves. The JC69 evolutionary model was used to generate  sequences of length 500. Three data sets were generated for each tree. We ran the annealed SMC, debiased adaptive annealed SMC (DASMC), and CSMC, respectively, for each data set three times with different random seeds.  In the annealed SMC, $\beta$ was set to 5, and $K$=100; in CSMC, the number of particles was set to 100,000. 

\begin{figure}[tp]
\begin{center}
\begin{tabular}{cc}
\rotatebox{90}{\quad \quad { Clock tree}} &
\includegraphics[width=0.9\textwidth]{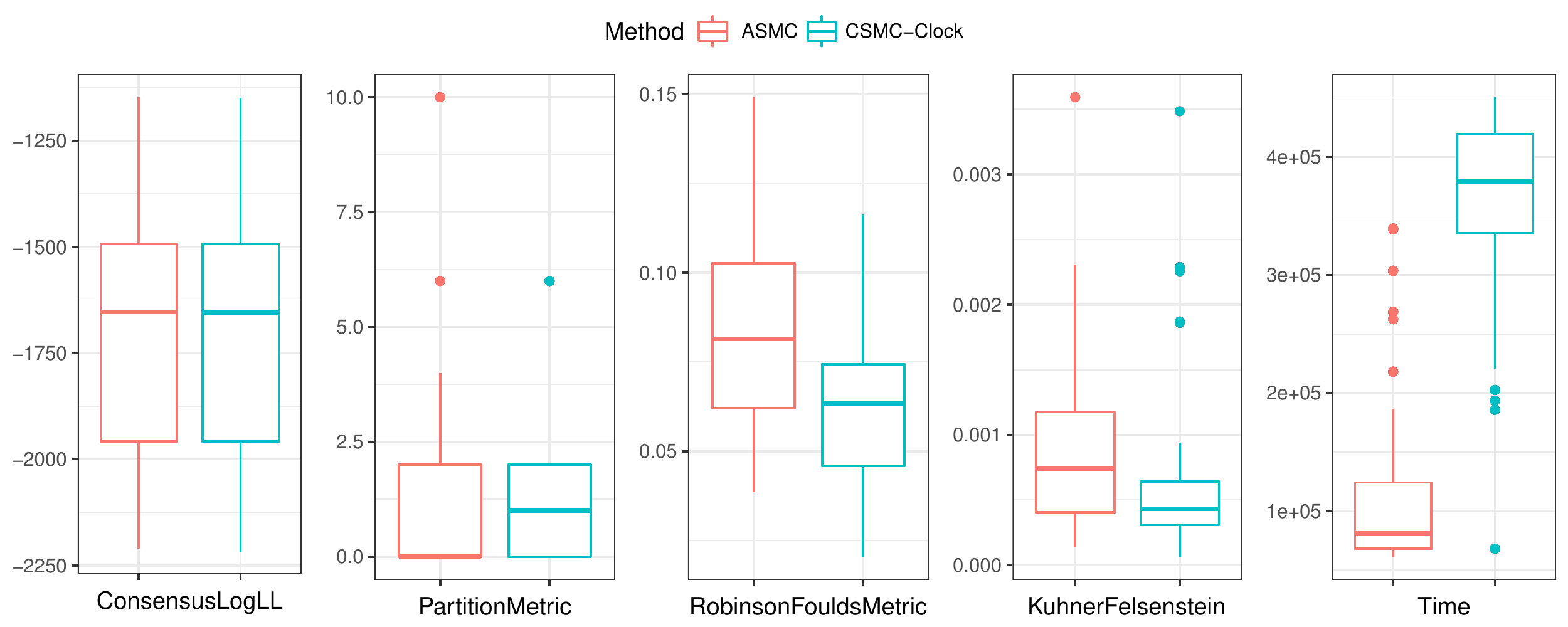}	\\
\rotatebox{90}{\quad { Relaxed clock tree}} &
\includegraphics[width=0.9\textwidth]{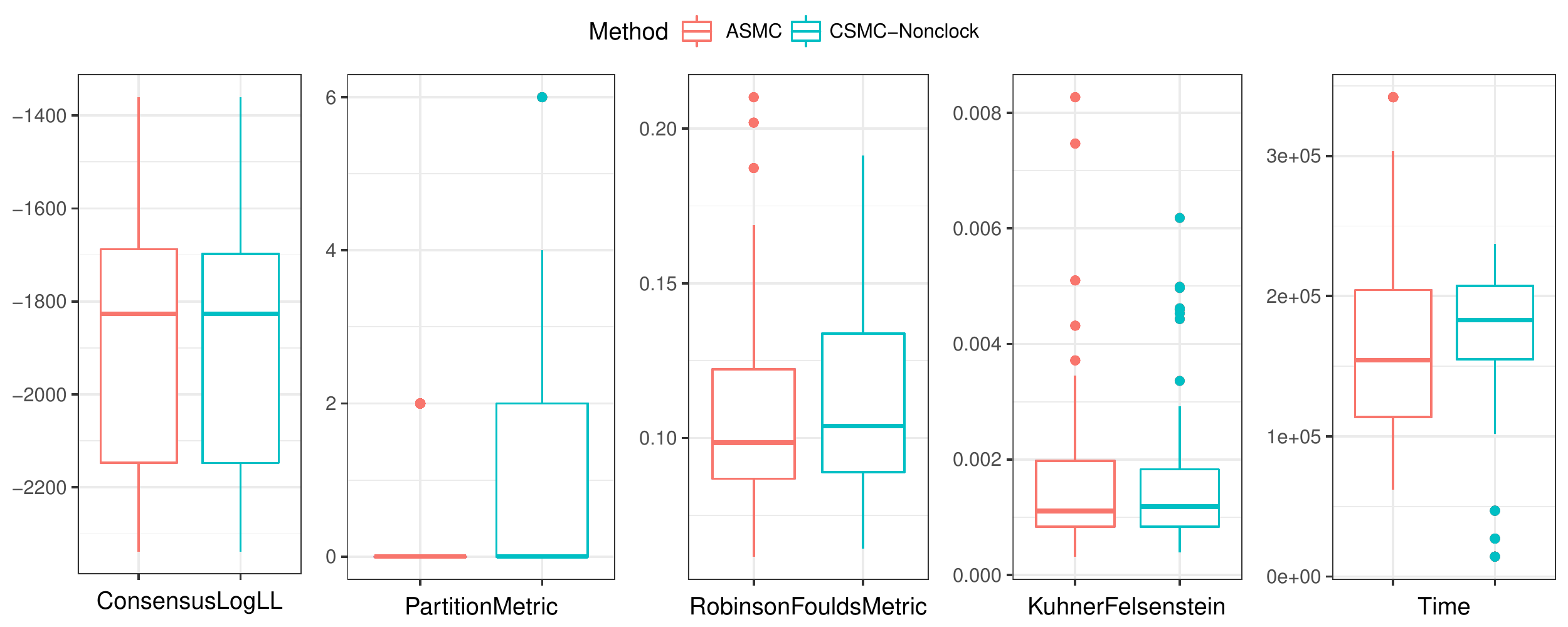}	\\
\rotatebox{90}{\quad { Nonclock tree}} &\includegraphics[width=0.9\textwidth]{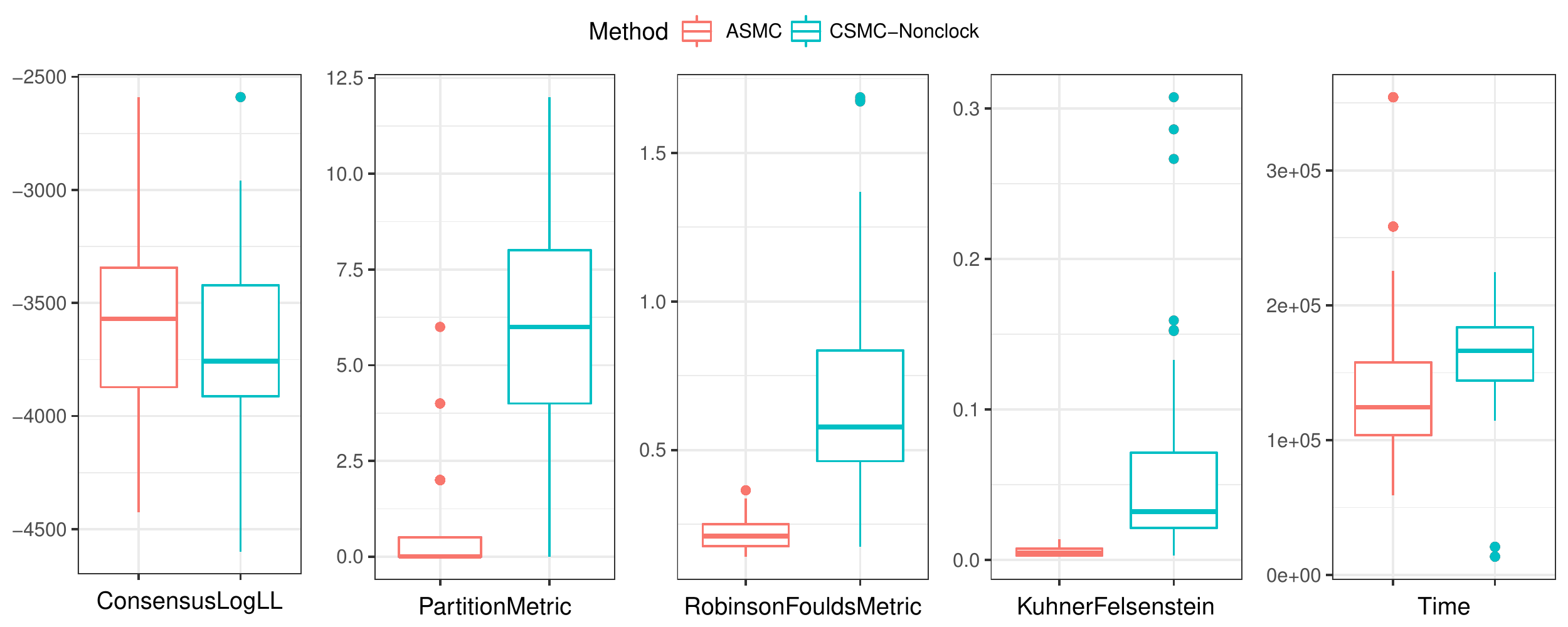}	\\
\end{tabular}
\caption{Comparison of adaptive SMC algorithms with CSMC for three types of simulated trees: clock, relaxed clock, nonclock (from top to bottom).}
\label{fig:treeDistanceASMCvsSMC}
\end{center}
\vspace{-0.2in}
\end{figure}

Figure \ref{fig:treeDistanceASMCvsSMC} shows the boxplots of log likelihood of the consensus trees, three tree distance metrics from the true trees, and computing time (in milliseconds) obtained from running the three algorithms for clock trees (top), relaxed clock trees (middle), and nonclock trees (bottom). Note that we ran CSMC for a longer time to favour this reference method. CSMC performs  well  for clock trees and relaxed clock trees, while the annealed SMC works for all of the three types of trees and clearly outperforms CSMC for non-clock trees.

\section{Real datasets}

We analyzed two difficult real data sets from TreeBASE: M336 and M1809 in Table 1 of \citet{Lakner01022008}. M336 contains DNA sequences of length $1949$ for $27$ species.  In M1809, there are $59$ species and the length of each DNA sequence is $1824$. We compared the marginal likelihood estimates, log-likelihood of the consensus tree, and tree distance metrics provided by ASMC and MrBayes (with the default setting) with the same computational budget. The reference trees used to compute tree distances are based on at least six independent long MrBayes parallel tempering runs provided by \cite{Lakner01022008}. Convergence to the posterior in these ``reference runs'' was established with high confidence in that previous work.  Note that the comparison handicaps ASMC as the set of tree moves in MrBayes is a superset of those used in ASMC. The evolutionary model we consider in real data analysis is the JC69 model.

\subsection{Dataset M336}

\begin{figure}[tp]
\begin{center}
\begin{tabular}{cc}
\includegraphics[width=1.0\textwidth]{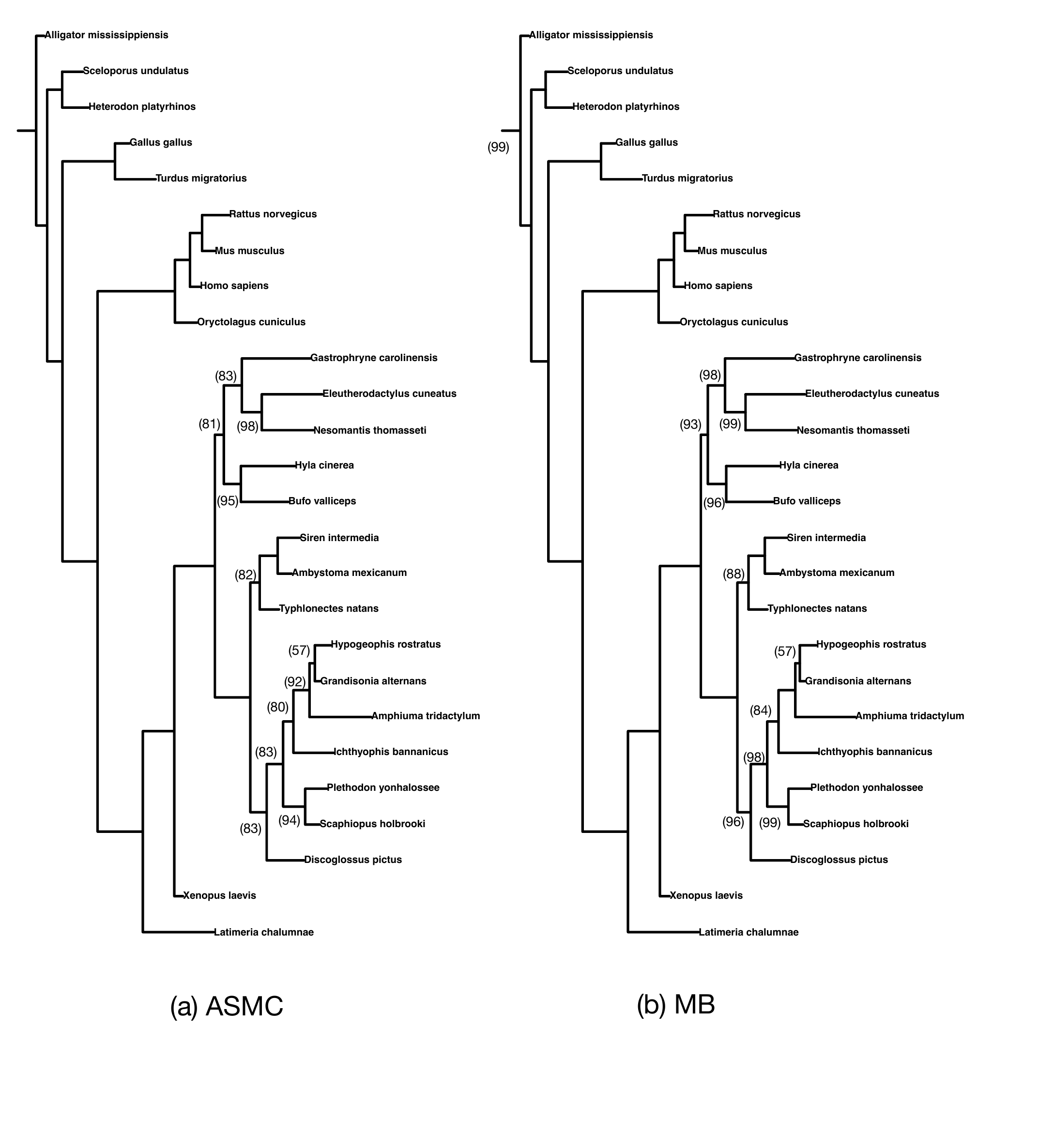}
\end{tabular}
\caption{ The majority-rule consensus trees for the M336 dataset estimated by (a) ASMC and (b) MrBayes. The numbers on the
trees represent the clade posterior probabilities (number 100 is omitted).}
\label{fig:cladePosterior}
\end{center}
\vspace{-0.2in}
\end{figure}

We used $K = 500$ and $\beta = 5.3$ for the ASMC algorithm. The log marginal likelihood estimated from ASMC is $-7103.73$, which is higher than the log marginal likelihood provided by MrBayes using Stepping Stone ($-7114.04$).
Table \ref{tab: realDataResults} displays the log-likelihood of the consensus tree and tree distance metrics provided by ASMC and MrBayes. In the table, $R$ represents the number of annealing parameters in ASMC and the total number of MCMC iterations in MrBayes respectively. The log-likelihood of the consensus tree estimated from ASMC is slightly lower than MrBayes. The RF and KF metrics estimated from MrBayes are slightly higher than ASMC. The majority rule consensus tree provided by ASMC and MrBayes are identical, and coincide with the reference tree. Figure \ref{fig:cladePosterior} displays the estimated majority-rule consensus trees and the clade posterior probabilities provided by ASMC and MrBayes. Most clades posterior probabilities provided by ASMC and MrBayes are close. ASMC provides lower posterior support for some clades, which is consistent with the hypothesized superior tree exploration provided by ASMC on a fixed budget.  

\begin{table}[htbp]
   \caption{Comparison of running ASMC and MrBayes for M336 from TreeBASE.}
   \label{tab: realDataResults}
   \centering
   \begin{tabular}{lllll}          
Method	 &	$R$  &	$K$	&	Metric  &	Value	 \\ \hline
ASMC   	& 	15706	&  500	&  ConsensusLogLL	 & -6892.16	 \\
&	15706	&  500	&	BestSampledLogLL   &	-6901.31	\\
& 15706	&  500		&	PartitionMetric	         &    0	                \\
& 15706 &  500  	&	RobinsonFouldsMetric &	0.01269	\\
& 15706	&  500	&	KuhnerFelsenstein	 & 5.55E-06	 \\
\hline
MrBayes		&		8.0E+06	& &	ConsensusLogLL &	-6889.52	\\
		&	8.0E+06		& &	PartitionMetric	&	0	 \\
		&	8.0E+06		&&	RobinsonFouldsMetric &	0.01832  \\
		&	8.0E+06		&&	KuhnerFelsenstein &	2.25E-5
\end{tabular}
\end{table}

\subsection{Dataset M1809}

We used $K = 1000$ and $\beta = 5$ for the ASMC algorithm. The log marginal likelihood estimated from ASMC is $-37,542.25$, the one estimated by MrBayes using Stepping Stone is $-37,335.73$. Table \ref{tab: real2} displays the tree metrics provided by ASMC and MrBayes. The log-likelihood of the consensus tree  provided by ASMC is higher than the one from MrBayes, and PM, RF, KF metrics estimated from ASMC are lower. 

\begin{table}[htbp]
   \caption{Comparison of running ASMC and MrBayes for M1809 from TreeBASE.}
   \label{tab: real2}
   \centering
   \begin{tabular}{lllll}          
Method	 &	$R$  &	$K$	&	Metric  &	Value	  \\ \hline
ASMC  	& 	17639	&  1000	&  ConsensusLogLL	 & -36,972.513 \\
&	17639	& 1000	&	BestSampledLogLL   &	-36,991.443	 \\
& 17639 &	1000	&	PartitionMetric	         &    2.0	                \\
& 17639 &   1000	&	RobinsonFouldsMetric &	0.13741	\\
& 17639 &	 1000	&	KuhnerFelsenstein	 & 3.95E-4 \\ \hline
MrBayes		&		1.76E+07	& &	ConsensusLogLL &	-36,996.13	\\
    	&		1.76E+07	& &	PartitionMetric	&	16.0	 \\
		&		1.76E+07	&&	RobinsonFouldsMetric &	0.513285  \\
		&		1.76E+07	&&	KuhnerFelsenstein &	0.01137
\end{tabular}
\end{table}

\section{Conclusion and discussion}

The annealed SMC algorithm discussed in this paper provides a simple but general framework for phylogenetic tree inference. Unlike previous SMC methods in phylogenetics, 
annealed SMC considers the same state space for all the intermediate distributions. 
As a consequence, many conventional Metropolis-Hastings tree moves used in the phylogenetic MCMC literature can be utilized as the basis of SMC proposal distributions. Since MCMC tree moves are available for a large class of trees, including non-clock as well as strict and relaxed clock models, the annealed SMC method is automatically applicable to a wide range of phylogenetic models.  It should also be relatively easy to incorporate the proposed ASMC into existing phylogenetic software packages that implement MCMC algorithms, such as MrBayes, RevBayes or BEAST.

The annealed SMC algorithm has two adaptive mechanisms, dynamic resampling and adaptive determination of annealing parameters, to make the algorithm efficient while requiring less tuning.  Dynamic resampling based on ESS is  a common practice in the SMC literature. Devising the annealing parameter sequence is a relatively newer practice \citep{del_moral_adaptive_2012}. The annealing parameter sequence can be determined dynamically based on the conditional ESS criterion. Since the particle weights of the current iteration only depend on the previous particles, there is negligible computational cost for finding annealing parameters.

The consistency of annealed SMC discussed in the theoretical results section holds when $K$ goes to infinity.  However, $K$ cannot in practice be made arbitrarily large as the memory requirements scale linearly in $K$. In contrast, increasing the number of intermediate distributions $R$ (in our adaptive algorithm, by increasing $\beta$) does not increase memory consumption. We conjecture that consistency for large $R$ but fixed $K$ also holds, in the sense of having the marginal distribution of each particle at the last iteration converging to the posterior distribution. 
We have explored the relative importance between $K$ and $R$ with fixed computational budgets using simulations.
These results suggest that increasing $R$ and $K$ improves the approximation at different rates, with increasing $R$ giving bigger bang to the buck. 
Assuming that our conjecture on the convergence in $R$ holds true, in the regime of very large $R$ and fixed $K$,  the particle population at the last iteration can be conceptualized as $K$ independent Monte Carlo samples from the true posterior distribution (in particular we conjecture that the weights will convergence to a uniform distribution and hence to naive Monte Carlo based on independent exact samples). We remind the reader that the power of independent exact Monte Carlo is that the variance does not depend on the dimensionality of the problem. Hence if $R$ is sufficiently large, a lower bound for $K$ can therefore be obtained by selecting $K$ large enough so that for independent samples $X_i$ with distribution $\pi$ the variance of the Monte Carlo average $(1/K) \sum_{k=1}^K f(X_i)$ is sufficiently small. Here is a concrete example: suppose we have a test function $f$ of interest, for example an indicator function on a fixed clade, with unknown posterior support $p = \int f(x) \pi(x) \ud x$. We should take $K$ large enough so that the Monte Carlo average will have a $95\%$ Monte Carlo confidence interval having a width of no more than say $\min\{p,1-p\}/10$. For $p \le 1/2$, this yields $K \ge (p/10)^{-2} (z^*)^2 \text{Var}_\pi f \approx 384 (1-p)/p$, where $z^* \approx 1.96$ is the $95\%$ critical value. For example, if the clade of interest is believed from a test run to be highly uncertain, $p \approx 1/2$, then in the large $R$ regime, at the very minimum $K=400$ particles should be used. The value of $K$ should also be sufficiently large to accommodate the number of parallel cores available, and also to ensure that the adaptive annealing scheme is stable (i.e. that a further increase in $K$ results in a qualitatively similar annealing schedule). See also \cite{olsson_numerically_2017} for more sophisticated schemes for estimating the Monte Carlo variance of SMC algorithms.

Importantly, annealed SMC provides an efficient way to estimate the marginal likelihood, which is still a challenging task in Bayesian phylogenetics. We have also reviewed other marginal likelihood estimation methods, including Stepping Stone and  Linked Importance Sampling. Our annealed SMC algorithm enjoys advantageous theoretical properties. The main property that justifies the use of the annealed SMC is the unbiasedness of its marginal likelihood estimate. In addition, the unbiasedness of the marginal likelihood estimate can be used to test implementation correctness of the algorithm.  Our simulation studies have shown that ASMC can give a similar marginal likelihood estimate as the one obtained from the ASMC with the same but deterministic annealing parameter sequence (debiased ASMC).  With the same computing budget, ASMC has been demonstrated to result in more accurate estimates.  Moreover, the ASMC algorithm requires less tuning than the other methods considered. Both LIS and SS need a predetermined annealing parameter sequence, which is often inconvenient to choose in practice.  ASMC leads to a more stable estimate for the marginal likelihood compared to the other methods considered.

MCMC moves often come with tuning parameters. For example, proposal distributions typically have a bandwidth parameter which needs to be tuned \citep{roberts_weak_1997}. To improve the performance of annealed SMC, it would be possible to use automatic tuning of proposal distributions within SMC algorithms, as proposed in  \cite{ZHOU2016TowardAutomaticModelComparison}.

A second future direction would be to investigate modifications in the specification of the sequence of intermediate distributions. For example, \cite{fan2010choosing} proposed an alternative to the prior distribution to replace $\pi_0$. The same choice could be used within our framework. In another direction, it may be possible to combine the construction of \cite{dinh2016online} with ours to handle online problems via standard moves: instead of integrating the new taxon with all of its sites un-annealed (which requires specialized proposals), it may be beneficial to anneal the newly introduced site.

 In terms of empirical comparisons, it would be interesting to expand the set of metrics, models and datasets used to compare the algorithms. For example, in addition to the tree distance metrics used in this article, geodesic tree distance  \citep{billera2001geometry} is also an important metric to compare distances between phylogenetic trees.
The GTP software \citep{owen2011fast} allows easy calculation of the geodesic tree distance.

We have investigated the subsampling SMC algorithm for ``tall data'' phylogenetic problems. The annealing parameters $\psi(s, \phi_{r})$ of the subsampling SMC is derived from the annealing parameter sequence $\phi_{r}$ of the adaptive ASMC without subsampling. This choice was made for implementation convenience, and there is no reason why the two optimal sequences of distributions should coincide. To improve the algorithm performance, one direction is therefore to design an adaptive scheme tailored to the subsampling version. One challenge is that pointwise evaluation of the adaptation function $g(\phi)$ is more expensive in the subsampling setup, with a cost that grows with $\phi$. Bayesian optimization might be useful in this context. 
Another use of the subsampling arises in situations where the sampling algorithm is not constructed using an accept-reject step.  For example, conjugate Gibbs sampling on an augmented target distribution \citep{lartillot2006conjugate} is used by PhyloBayes \citep{lartillot_phylobayes_2009} to efficiently sample the evolutionary model parameters. It is not clear how the conjugate Gibbs sampling step can be modified to accommodate the annealed distribution in Equation~(\ref{eqn:piTophi}) for $\phi < 1$. On the other hand, conjugate sampling is directly applicable to intermediate distributions that consist in taking subsets of sites. This sequence of distributions could be used to handle conjugate Gibbs sampling not only in annealed SMC but also in the context of parallel tempering or any other sequence of measure based method. 
One last line of work is to combine control variates to annealed SMC to reduce the variance of the likelihood estimator, a general strategy that has been very successful in other subsampling work \citep{bardenet2017markov}.

\section{FUNDING}
This research was supported by Discovery Grants of A.  Bouchard-C\^{o}t\'{e} and L. Wang from the National Science and Engineering Research Council and a Canadian Statistical Sciences Institute Collaborative Research Team Project. Research was enabled in part by support provided by WestGrid (\url{www.westgrid.ca}) and Compute Canada (\url{www.computecanada.ca}). 

\section{ACKNOWLEDGMENTS}

We would like to thank the reviewers and editor for the constructive feedback, as well as Arnaud Doucet and Fredrik Ronquist for helpful discussions. Part of this work was done while visiting the Department of Statistics at Oxford.

\bibliographystyle{sysbio}

\section{APPENDIX 1}

\subsection{Construction of Intermediate Distributions for Subsampling SMC }
\label{sec:subsampling}

Let us decompose the unnormalized posterior distribution as
\[
\gamma_{1}(x) = p(x)\prod_{s = 1}^{\# S}p(y_{s}|x),
\]
where $x$ refers to the phylogenetic tree and evolutionary parameter of interest, $s$ is an index for one batch of sites from a biological sequence, and \#$S$ represents the total number of batches. Each batch contains one or more sites of the biological sequence; we denote the number of sites in each batch by $b_s$. 

Consider the annealing parameter sequence $0 = \phi_{0} < \phi_{1} <\dots  < \phi_{R} = 1$. We define the  sequence  of 
intermediate distributions for subsampling as follows:
\[
\gamma_{\phi_{r}}(x) = p(x)\prod_{s = 1}^{\# S}p(y_{s}|x)^{\psi(s, \phi_{r})},
\]
where 
\[ \psi(s, \phi_{r}) =
  \begin{cases}
    1       & \quad \text{if } \phi_{r} \geq  s/\#S,\\
    0  & \quad \text{if } \phi_{r} \leq  (s-1)/\#S, \\
   \#S\cdot\phi_{r} -  (s-1)   & \quad \text{otherwise. }
  \end{cases}
\]
The subsampling SMC algorithm is a more general version of the annealed SMC algorithm. If we define $\#S = 1$ in $\gamma_{1}(x)$, then the sequence of intermediate distributions of subsampling SMC is exactly the same as the intermediate distributions of the annealed SMC. In this case, the computational cost of subsampling SMC is exactly the same as the annealed SMC. Another extreme case is that $\#S = n$, in which case we sequentially incorporate the sites of sequence one by one.

\section{APPENDIX 2}
\label{sec:app2}

\subsection{Theoretical Foundations of Annealed SMC}

In this section, we review the construction of \cite{del2006sequential}, which is the basis for our work. See also \cite{LiangliangWang2015} for a similar construction tailored to a phylogenetic setup.

The corresponding sequence of unnormalized  distributions 
are denoted by  $\{\gamma_r\}_{1,\ldots, R}$. 
The annealed SMC can be obtained by defining 
an auxiliary sequence of distributions that admit the distribution of interest,
$\pi_r(x_r)$, as the marginal of the latest iteration 
$$\tilde{\pi}_r(\xbold_r)=\pi_r(x_r)\prod_{j=1}^{r-1}L_j(x_{j+1},x_j),$$
where $L_j(x_{j+1},x_j)$ is an auxiliary ``backward'' Markov kernel with $\int L_j(x_{j+1},x_j) \ud x_j = 1$. We never sample from $L_j$, rather its role is to allow us to derive weight updates that yield a valid SMC algorithm.

The idea is then to apply standard SMC (i.e. SMC for product spaces such as state space models) to this auxiliary sequence of distributions, $\tilde{\pi}_1, \tilde{\pi}_2, \dots, \tilde{\pi}_R$.  
The resulting sampler has a weight update given by
\begin{align*}w(x_{r-1},x_{r}) &\propto \frac{\tilde{\pi}_r(\xbold_r)}{\tilde{\pi}_r(\xbold_{r-1})} \frac{1}{{K}_{r}(x_{r-1},x_{r})}  \\ &= \frac{\pi_r(x_{r})L_{r-1}(x_{r},x_{r-1})}{\pi_{r-1}(x_{r-1}){K}_{r}(x_{r-1},x_{r})},
\end{align*}
which is different from the one in a
standard SMC. 
 
When $K_r$ satisfies global balance with respect to $\pi_r$, a convenient backward  Markov kernel that allows an easy evaluation of the importance weight is
\begin{align*}\label{eqn:backwardKernel}
L_{r-1}(x_r,x_{r-1})=\frac{\pi_r(x_{r-1}){K}_r(x_{r-1},x_r)}{\pi_r(x_r)}.
\end{align*}
This choice is a properly normalized backward kernel, $\int L_{r-1}(x_r,x_{r-1}) \ud x_{r-1} = 1$: this follows from the assumption that $K_r$ satisfies global balance with respect to $\pi_r$.  
With this backward kernel, the incremental importance weight becomes
\begin{eqnarray*}
	w(x_{r-1},x_{r}) &=& \frac{\gamma_r(x_{r})}{\gamma_{r-1}(x_{r-1})}\cdot
	\frac{L_{r-1}(x_{r}, x_{r-1})}{{K}_{r}( x_{r-1}, x_{r})} \\ &=& 
	\frac{\gamma_r(x_{r})}{\gamma_{r-1}(x_{r-1})} \cdot
	\frac{\pi_r(x_{r-1}){K}_r(x_{r-1}, x_r)}{\pi_r(x_r)} \cdot \frac{1}{K_{r}(x_{r-1}, x_{r})} \\ &=& 
	\frac{\gamma_r(x_{r-1})}{\gamma_{r-1}(x_{r-1})}.  
\end{eqnarray*}

\subsection{General Estimates of Marginal Likelihood} 

In Section \emph{Basic Annealed SMC Algorithm} we describe the estimator for marginal likelihood in a simplified setting, i.e. without adaptation. Here we describe the marginal likelihood estimator in full generality.

Recall that we denote the marginal likelihood by $Z$ for simplicity.  With a slight abuse of notation, we use $K_{r}(x_{r-1},\cdot)$ to denote the proposal distribution for $x_{r}$ in this section. 

Let us start by rewriting the normalization constant of the first intermediate distribution as 
 \begin{eqnarray*}
 Z_1=\int \frac{\gamma_1(x_1)}{K_1(x_1)}K_1(x_1)dx_1=\int
 w_1(x_1) K_1(x_1)dx_1,
 \end{eqnarray*}
 where $K_1(\cdot)$ is the proposal distribution for $x_1$. 
 
 Correspondingly, an estimate of $Z_1$ is
 $$Z_{1,K}=\frac{1}{K}\sum_{k=1}^Kw_{1,k}.$$
Similarly, we can rewrite 
the ratio of the normalization constants of two intermediate distributions as 
\begin{eqnarray*}
\frac{Z_r}{Z_{r-1}}&=&\frac{\int \gamma_r(x_r)dx_r}{Z_{r-1}}
=\frac{\int
\gamma_r(x_r)dx_r}{\gamma_{r-1}(x_{r-1})/\pi_{r-1}(x_{r-1})}
\\
&=& \int
\frac{\gamma_r(x_r)}{\gamma_{r-1}(x_{r-1})}\pi_{r-1}(x_{r-1})dx_r\\
&=& \int
\frac{\gamma_r(x_r)}{\gamma_{r-1}(x_{r-1})K_r(x_{r-1},
x_{r})}\pi_{r-1}(x_{r-1})K_r(x_{r-1},
x_{r})dx_r\\
&=& \int
w_r(x_r)\pi_{r-1}(x_{r-1})K_r(x_{r-1},
x_{r})dx_r.
\end{eqnarray*}
Straightforwardly, an estimate of ${Z_r}/{Z_{r-1}}$ is provided by 
 \begin{eqnarray*}
 \widehat{\frac{Z_r}{Z_{r-1}}}= \frac{1}{K}\sum_{k=1}^K w_{r,k}.
 \end{eqnarray*}
Since the estimate of the  marginal likelihood can be rewritten as  
  \begin{eqnarray*}
  Z\equiv Z_R=Z_1\prod_{r=2}^R\frac{Z_r}{Z_{r-1}}, 
  \end{eqnarray*}  
an estimate of the marginal likelihood $Z$ is   
\begin{eqnarray}
  \hat{Z}_{R,K}=\prod_{r=1}^R\left(\frac{1}{K}\sum_{k=1}^K
  w_{r,k}\right) =\prod_{r=1}^R\left(\frac{1}{K}\sum_{k=1}^K  \{p(y|x_{r-1,k})\}^{\phi_r -\phi_{r-1}} \right),
  \end{eqnarray}
  which can be obtained from an {SMC} algorithm readily.
If resampling is not conducted at each iteration $r$, an alternative  form is provided by
\begin{eqnarray}\label{eqn:margin:zHat2}
  \widehat{Z_{R,K}} =\prod_{j=1}^{t_{R-1}+1}\left(\sum_{k=1}^K W_{n_{j-1}, k}\prod_{m = n_{j-1}+1}^{n_{j}}  \{p(y|x_{m-1,k})\}^{\phi_m -\phi_{m-1}} \right),
  \end{eqnarray}
 where $n_{j}$ is the SMC iteration index at which we do the $j$th resampling, $t_{R-1}$ is the number of resampling steps between $1$ and $R-1$.
 
\subsection{Unbiasedness, Consistency and Central Limit Theorem for Annealed SMC} 
Here we provide more information on the theoretical properties discussed in Section \emph{Properties of Annealed Sequential Monte Carlo}. 
 
\noindent {\bf Theorem 1} ({\it Unbiasedness}): For fixed $ 0=\phi_{0} < \phi_{1}<\cdots <\phi_{R} = 1$, $ \hat{Z}_{R,K}$ is an unbiased estimate of $Z$, 
\[
 \E(\hat{Z}_{R,K}) = Z.
\]
This result is well known in the literature, although many statements of the result are specialized to SMC for state space models \citep{Doucet11tutorial}. 
The results in {\it Theorem 7.4.2} of \cite{DelMoral2004} provide a very general set of conditions which includes the annealed SMC algorithm presented here. However, the theoretical framework in \cite{DelMoral2004} being very general and abstract, we outline below an alternative line of argument to establish unbiasedness of phylogenetic annealed SMC. 

First, by the construction reviewed in Section \emph{Theoretical Foundations of Annealed SMC}, we can transform the sequence of distributions on a fixed state space, $\pi_r(x_r)$, into a sequence of augmented distributions $\tilde{\pi}_r(\xbold_r)$ on a product space admitting $\pi_r(x_r)$ as a marginal. We now apply Theorem~2 of \cite{Andrieu2010} with the distribution $\pi_n(x_{1:n})$ in this reference set to $\tilde{\pi}_r(\xbold_r)$ in our notation. To be able to use Theorem~2, we only need to establish the ``minimum assumptions'' 1 and 2 in \cite{Andrieu2010}. Assumption 1 is satisfied by the fact that valid MCMC proposals are guaranteed to be such that $q(x, x') > 0 \Longleftrightarrow q(x', x) > 0$. Assumption 2 holds since we use multinomial resampling. Next, since the conditions of Theorem~2 of \cite{Andrieu2010} hold, we have the following result from the proof of Theorem~2 in Appendix~B1 of \cite{Andrieu2010}:
\[
\frac{\tilde \pi^N(k, \boldsymbol{\bar x_1}, \dots, \boldsymbol{\bar x_P}, \boldsymbol{a_1}, \dots, \boldsymbol{a_{P-1}})}{q^N(k, \boldsymbol{\bar x_1}, \dots, \boldsymbol{\bar x_P}, \boldsymbol{a_1}, \dots, \boldsymbol{a_{P-1}})} = \frac{\hat Z^N(\boldsymbol{\bar x_1}, \dots, \boldsymbol{\bar x_P})}{Z},
\]
and hence
\[
Z \; \tilde \pi^N(k, \boldsymbol{\bar x_1}, \dots, \boldsymbol{\bar x_P}, \boldsymbol{a_1}, \dots, \boldsymbol{a_{P-1}}) = \hat Z^N(\boldsymbol{\bar x_1}, \dots, \boldsymbol{\bar x_P}) \; q^N(k, \boldsymbol{\bar x_1}, \dots, \boldsymbol{\bar x_P}, \boldsymbol{a_1}, \dots, \boldsymbol{a_{P-1}}).
\]
Now taking the integral on all variables $k, \boldsymbol{\bar x_1}, \dots, \boldsymbol{\bar x_P}, \boldsymbol{a_1}, \dots, \boldsymbol{a_{P-1}}$ with respect to the reference measure $\mu$ associated to $\tilde \pi^N$, we obtain:
\begin{align*}
&Z \int \tilde \pi^N(k, \boldsymbol{\bar x_1}, \dots, \boldsymbol{\bar x_P}, \boldsymbol{a_1}, \dots, \boldsymbol{a_{P-1}}) \ud \mu(k, \boldsymbol{\bar x_1}, \dots, \boldsymbol{\bar x_P}, \boldsymbol{a_1}, \dots, \boldsymbol{a_{P-1}}) \\ 
&= \int \hat Z^N(\boldsymbol{\bar x_1}, \dots, \boldsymbol{\bar x_P}) q^N(k, \boldsymbol{\bar x_1}, \dots, \boldsymbol{\bar x_P}, \boldsymbol{a_1}, \dots, \boldsymbol{a_{P-1}})  \ud \mu(k, \boldsymbol{\bar x_1}, \dots, \boldsymbol{\bar x_P}, \boldsymbol{a_1}, \dots, \boldsymbol{a_{P-1}}).
\end{align*}
The left-hand side is just $Z$ since $\tilde \pi^N$ is a density with respect to $\mu$. For the right-hand side, note that $q^N$ is the law of the full set of states produced by the particle filter, hence the right-hand side is just $\E[\hat Z_{R,K}]$ in our notation. This concludes the proof.

Next, we discuss consistency results. In the SMC literature, they are generally available both in the $L^{2}$ convergence and almost sure convergence flavours. We cover the $L^{2}$ case here and refer to \cite{del2006sequential} for almost sure consistency results. 
 
\noindent {\bf Theorem 2} ({\it Consistency}): Assume there is a constant $C$ such that $|f| \le C$ and $w_{r,k} \le C$ almost surely.  For a fixed $\phi_{r}$ $(r = 1,\ldots, R)$, the annealed SMC algorithm provides asymptotically consistent estimates:
\[
 \sum_{k=1}^K W_{r,k} f(x_{r,k})  \to \int \pi_r(x) f(x) \ud x ~~\text{as}~~K\to \infty,
\]
where the convergence holds in the $L^{2}$ norm sense.

The result can be deduced from Proposition~5 in \cite{LiangliangWang2015} as follows. Assumption~3 in \cite{LiangliangWang2015} holds since MCMC proposals satisfy $q(x, x') > 0 \Longleftrightarrow q(x', x)>0$. Assumption~4 holds since the support of the prior coincides with support of the posterior. 

Note that the assumption that the weights are bounded is not valid for general tree spaces. However, if the branch lengths are assumed to be bounded then the space is compact and the assumption therefore holds in that setting.

Finally, we turn to the central limit theorem. 

\noindent {\bf Theorem 3} ({\it Central Limit Theorem}): Under the integrability conditions given in Theorem $1$ of \cite{chopin2004central}, or \cite{DelMoral2004}, section $9.4$, pages $300-306$,
\[
K^{1/2}\bigg[ \sum_{k=1}^K W_{r,k} f(x_{r,k})  - \int \pi_r(x) f(x) \ud x\bigg]\to N(0, \sigma_{r}^{2}(f)) ~~\text{as}~~K\to \infty,
\]
where the convergence is in distribution. The form of asymptotic variance $\sigma_{r}^{2}(f)$ depends on the resampling scheme, the Markov kernel $K_{r}$ and the artificial backward kernel $L_{r}$. We refer readers to \cite{del2006sequential} for details of this asymptotic variance.

\section{APPENDIX 3}  

\subsection{MCMC Proposals for Bayesian Phylogenetics}
\label{sec:MCMCproposal}

In this paper,  we used the proposals $q_r^i$  defined as follow:
\begin{enumerate}
  \item  $q_r^1$: the \emph{ multiplicative branch proposal}. 
This proposal picks one edge at random and multiply its current value by a  random number
distributed uniformly in $[1/a , a]$ for some
fixed parameter $a > 1$ (controlling how bold the move is)
\citep{Lakner01022008}.
  \item  $q_r^2$: the \emph{global multiplicative branch proposal} that proposes
  all the branch lengths by applying the above  multiplicative branch proposal to each
  branch.
  \item  $q_r^3$: the \emph{stochastic NNI 
  proposal}. 
  We consider the nearest neighbor interchange (NNI) \citep{Jow2002} to propose  a new
tree topology.  
 \item  $q_r^4$: the \emph{stochastic NNI proposal with resampling the
   edge} that uses the above NNI 
  proposal in (3) and the  multiplicative branch proposal in (1) for the edge under
  consideration. 
 \item  $q_r^5$: the \emph{Subtree Prune and Regraft (SPR) move} that selects and removes a subtree from the main tree and reinserts it elsewhere on the main tree to create a new tree.
\end{enumerate} 
Note that here we only describe the MCMC kernels for phylogenetic trees. To sample evolutionary parameters
$\theta$, one can use simple proposals such as symmetric Gaussian distributions, or more complex ones, see for example \cite{zhao_bayesian_2016}.

\section{APPENDIX 4}  

\subsection{Comparison of Resampling Strategies}

\begin{figure}[b]
    \centering
        \includegraphics[width=0.9\textwidth]{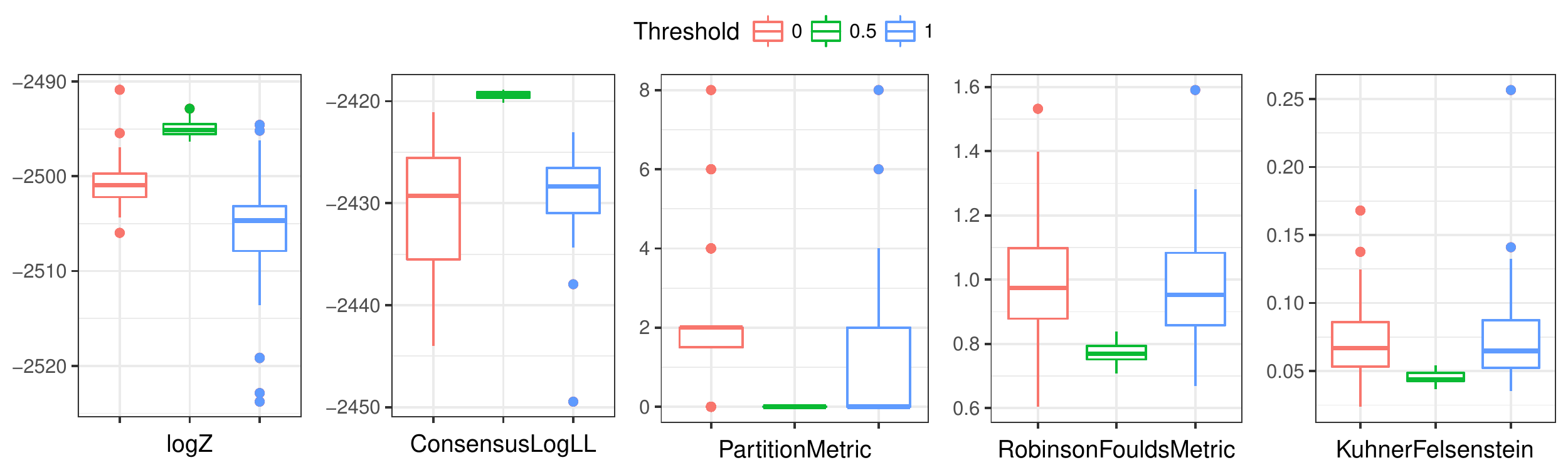}
        \caption{Comparison of three resampling thresholds, $0$, $0.5$ and $1$. }
        \label{fig:ResamplingThresholds}
\end{figure}

In this section we compare the performance of adaptive annealed SMC with three different resampling thresholds. The first resampling threshold is $\epsilon_{1} = 0$. In this case, the particles are never resampled. The second resampling threshold is $\epsilon_{2} = 1$, in which case resampling is triggered at every iteration. The third resampling threshold is $\epsilon_{3} = 0.5$. The resampling method we used in all our experiments was the multinomial resampling scheme. 
We simulated one unrooted tree of $15$ taxa, and  generated one data set of DNA sequences of length $200$. The tree simulation setup was the same as in Section \emph{Simulation Studies}. We ran adaptive annealed SMC algorithm  $20$ times with the three resampling thresholds described above. We used $rCESS_{r} = 0.99999$ and $K = 100$. Figure \ref{fig:ResamplingThresholds} demonstrates the 
advantage of resampling triggered by a threshold of $\epsilon_3 = 0.5$ over the other two choices by displaying
the marginal likelihood estimates, log likelihood of the consensus  tree and tree metrics provided by adaptive annealed SMC using three different $\epsilon$'s.  The log  marginal likelihood estimate and log likelihood of the consensus tree provided by adaptive annealed SMC using $\epsilon_{3} = 0.5$ are higher and admit smaller variation. The PF, RF and KF metrics provided by adaptive annealed SMC using $\epsilon_{3} = 0.5$ are lowest.  Therefore  the threshold 0.5 has been used in the rest of the paper. 

\section{APPENDIX 5}  
\subsection{Review of Particle Degeneracy Measures}
The two adaptive schemes in ASMC, adaptively conducting resampling and the automatically construction of the annealing parameter sequence,
rely on being able to assess the quality of a particle approximation. For completeness, we provide some background in this section on the classical notation of Effective Sample Size (ESS) and of  conditional ESS (CESS), a recent generalization which we use here \citep{ZHOU2016TowardAutomaticModelComparison}. The notion of ESS in the context of importance sampling (IS) or SMC is distinct from the notion of ESS in the context of MCMC. The two are related in the sense of expressing a variance inflation compared to an idealized Monte Carlo scheme but they differ in the details. We will assume from now on that ESS refers to the SMC context.

We will also use a slight variation of the derivation of ESS and CESS where the measures obtained are normalized to be between zero and one (some hyper-parameters of the adaptive algorithms are easier to express in this fashion).  We use the terminology relative (conditional) ESS to avoid confusion.

The fundamental motivation of (relative and/or conditional) ESS stems from the analysis of the error of Monte Carlo estimators. Recall that for a given function of interest $f$ (think for example of $f$ being an indicator function on a clade), 
\[
I = \int \pi_r(\ud x) f(x) \approx  \sum_{k=1}^K W_{r,k} f(x_{r,k}) =: \hat I.
\]
The quantity on the right hand side is a random variable (with respect to the randomness of the SMC algorithm), $\hat I$, and we can think about it as an estimator of the deterministic quantity on the left hand side, $I$. Moreover the right hand side is a real-valued random variable so we can define its mean square error, which can be further decomposed as a variance term and a squared bias term. For SMC algorithms, the variance term dominates as the number of particles goes to infinity \citep{DelMoral2004}. For this reason, we are interested in estimates of the variance of $\hat I$ across SMC random seeds, $\varsmc  [\hat I]$. 
However, the variance of $\hat I$ depends on the choice of function $f$, which is problem dependent, and we would like to remove this dependency. The first step is to consider a notion of relative variance, comparing to the variance we would obtain from a basic Monte Carlo scheme $\hat I^*$ relying on iid exact samples $x_1^\star, \dots, x_K^\star \sim \pi$, $\varbmc[ \hat I^*] = \text{Var}[\frac{1}{K} \sum_k f(x_k^\star)]$.
To make further progress, we will make approximations of the ratio $\varbmc[\hat I^*] / \varsmc[\hat I]$.

To understand these approximations, let us start with a simplified version of Algorithm~\ref{algo:adapt}, where the function NextAnnealingParameter returns the value 1.0. In this setting, no resampling occurs, and the algorithm reduces to an importance sampling algorithm (more specifically, it reduces to a single iteration of the Annealed Importance Sampling (AIS) algorithm \citep{neal_annealed_1998}). Importance sampling is easier to analyze since the individual particles are independent and identically distributed, allowing us to summarize the behaviour based on one particle, say $k=1$.  If we assume further that (A1) $\gamma_i = \pi_i$, i.e. that the normalization constant is one, then a classical argument by \cite{kong_ess_1992} based on the Delta method yields
\begin{align*}
\frac{\varbmc[\hat I^*]}{\varsmc[\hat I]} = \frac{\varbmc[\hat I^*]}{\text{Var}_{\pi_0}[\hat I]} &\approx \frac{1}{1 + \text{Var}_{\pi_0}[\tilde w_{1,1}]}, \\
&= \frac{1}{\E_{\pi_0}\left[ (\tilde w_{1,1})^2 \right]},
\end{align*}
where we used the fact that in this simple setting the distribution of one proposed particle is just $\pi_0$, so $\varsmc[\cdot] = \text{Var}_{\pi_0}[\cdot]$ and in the last line,
\[ \E_{\pi_0}[ \tilde w_{1,1} ] = \int \pi_0(x_{0,1}) \frac{\pi_1(x_{0,1})}{\pi_0(x_{0,1})} \ud x_{0,1} = 1. \] 

In general, assumption (A1) does not hold, i.e. the normalization constant is not one, so for a general one-step AIS algorithm we get instead the approximation:
\begin{align*} 
\frac{\varbmc[\hat I^*]}{\varsmc[\hat I]} &\approx \left(\E_{\pi_0}\left[ \left(\frac{\pi_1(x_{0,1})}{\pi_0(x_{0,1})}\right)^2 \right] \right)^{-1} \\
&= \left( \E_{\pi_0}\left[ \left(\frac{\gamma_1(x_{0,1})/Z_1}{\gamma_0(x_{0,1})/Z_0}\right)^2 \right] \right)^{-1} \\
&= \left( \frac{Z_1}{Z_0} \right)^2 \Big\slash\; \E_{\pi_0}\left[ \left(\frac{\gamma_1}{\gamma_0}(x_{0,1})\right)^2 \right].
\end{align*}

Generalizing the notation of this section into a general SMC setup, $\pi_1$ here plays the role of the current iteration, and $\pi_0$, of the previous iteration. However, since $\pi_0$ is not known in this case, we plug-in a particle approximation $\hat \pi_0 = \sum_{k=1}^K W_{0,k} \delta_{x_{0,k}}$ to get:
\[  \frac{\varbmc[\hat I^*]}{\varsmc[\hat I]} \approx \left( \frac{Z_1}{Z_0} \right)^2 \Big\slash\; \E_{\hat \pi_0}\left[ \left(\frac{\gamma_1}{\gamma_0}(x_{0,1})\right)^2 \right] = \left( \frac{Z_1}{Z_0} \right)^2 \Big\slash  \sum_{k=1}^K W_{0,k} \left( \frac{\gamma_1}{\gamma_0}(x_{0,k})  \right)^2  . \]
The effect of this additional approximation is that it makes our estimator over-optimistic, by ignoring the error of the approximation $\hat \pi_0$ of $\pi_0$. It is nonetheless a useful tool to assess the degradation of performance over a small number of  SMC iterations.

Finally, since the ratio of normalization constants is also unknown, we also need to estimate it. Based on a particle approximation of Equation~(\ref{eq:ratio-estimator}), we obtain:
\begin{align*}
\frac{\varbmc[\hat I^*]}{\varsmc[\hat I]} &\approx 
	\left( 
		\sum_{k=1}^K W_{0,k} \frac{\gamma_1}{\gamma_0}(x_{0,k})  
	\right)^2 
	\Big\slash
	\sum_{k=1}^K W_{0,k} 
	\left( 
		\frac{\gamma_1}{\gamma_0}(x_{0,k})  
	\right)^2. 
\end{align*}
 
This quantity is called the relative conditional ESS (rCESS), Equation~(\ref{eq:rcess-def}).
Having a high rCESS value is a necessary but not sufficient condition for a good SMC approximation. If it is low during some SMC iteration, especially an iteration close to the final iteration, then with high probability most of the particles will have very small or zero weights, which will lead to a collapse of the quality of the annealed SMC algorithm.

\subsection{Comparison of Relative ESS (rESS) and Relative CESS (rCESS)}

\begin{figure}
     \centering
     \begin{subfigure}[b]{0.4\textwidth}
         \includegraphics[width=\textwidth]{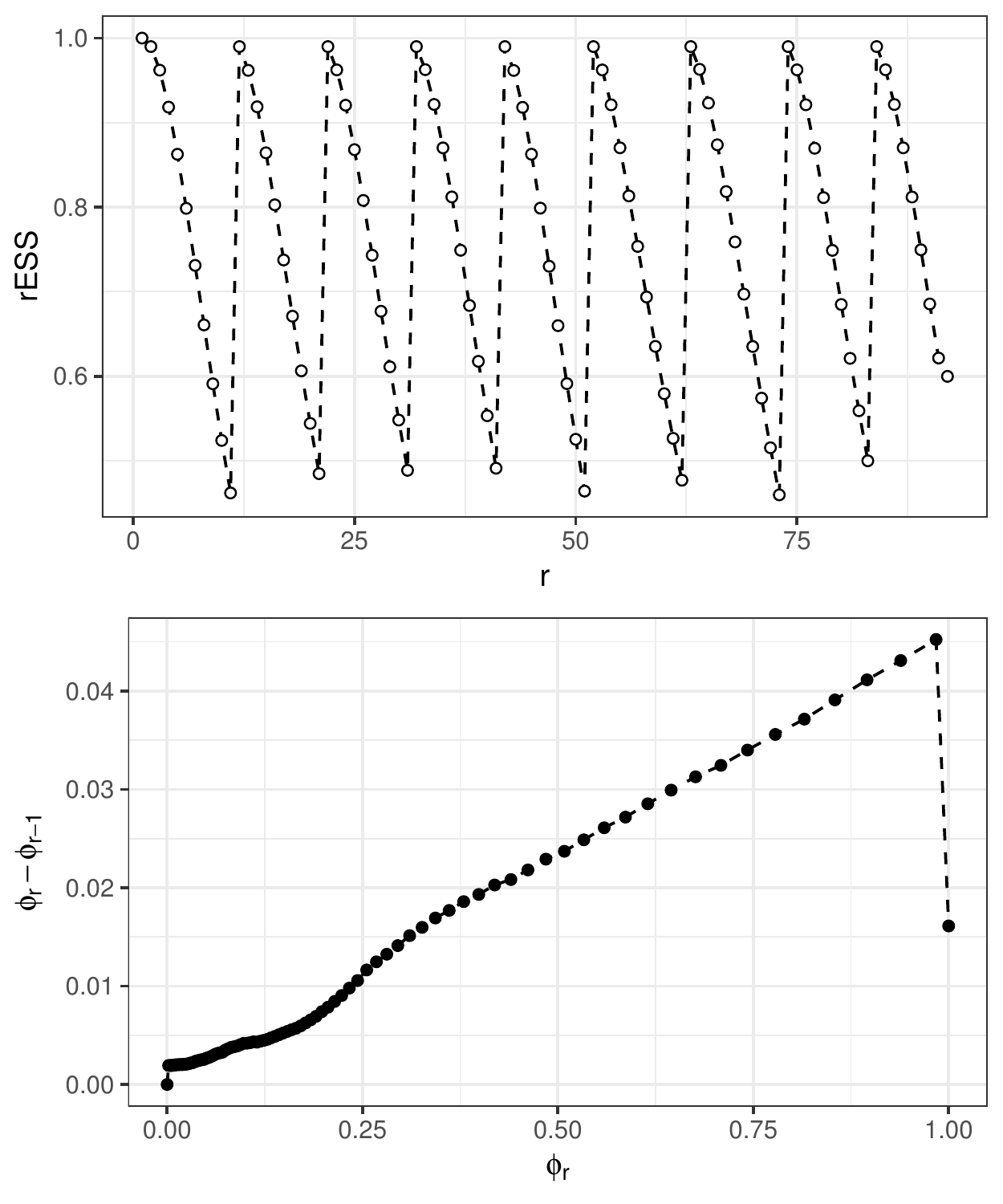}
         \caption{rCESS}
         \label{fig:CESS}
     \end{subfigure}
     \begin{subfigure}[b]{0.4\textwidth}
         \includegraphics[width=\textwidth]{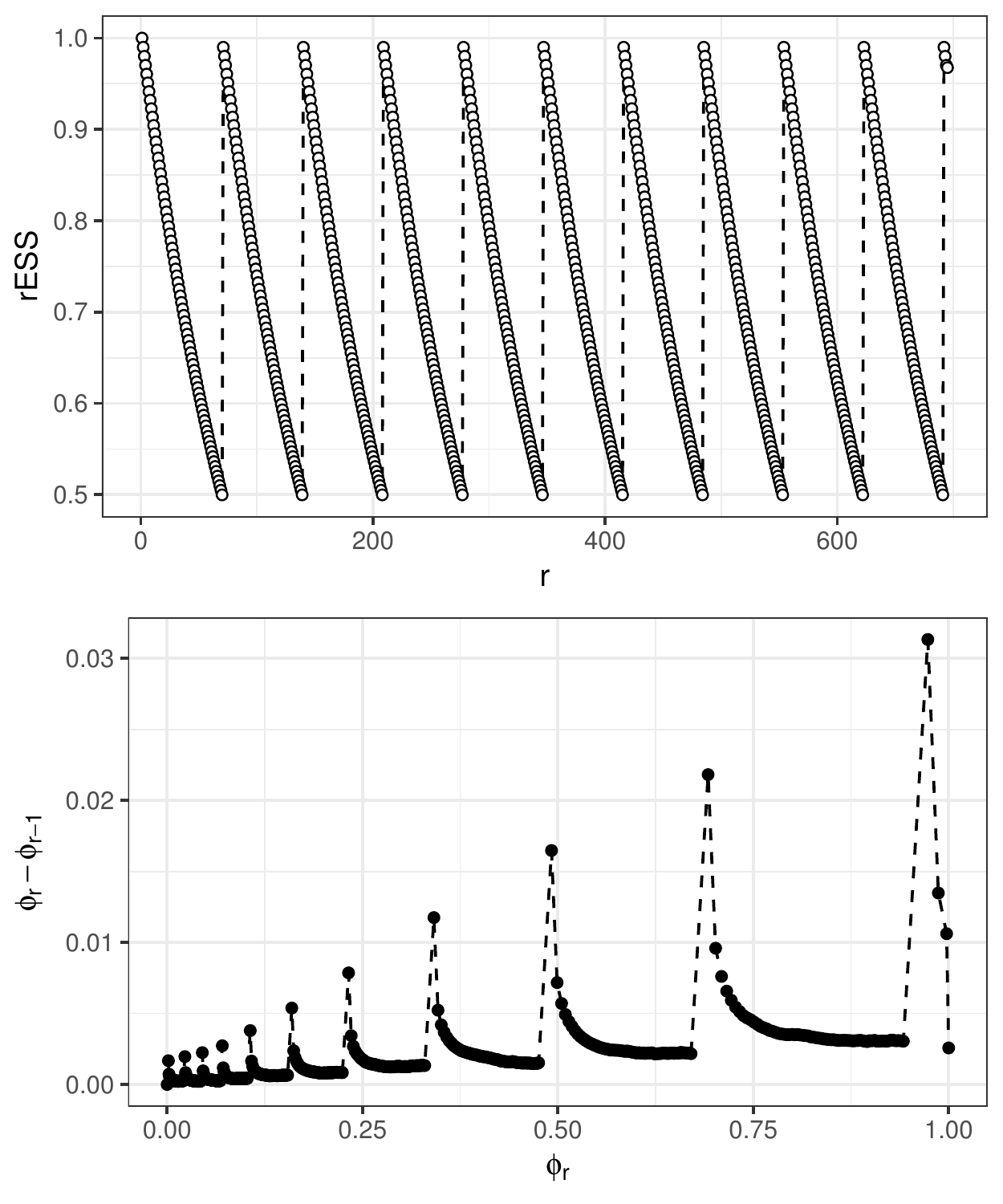}
        \caption{rESS}
         \label{fig:ESS}
     \end{subfigure}
     \caption{Comparison of rCESS (left) and rESS (right) in terms of rESS as a function of $r$ (top) and $\phi_{r} - \phi_{r-1}$ as a function  of $\phi_{r}$ (bottom).} \label{fig:CESS and ESS}
 \end{figure}

In earlier work on adaptive SMC methods, the function NextAnnealingParameter was implemented using a different criterion based on rESS instead of rCESS. Later, \cite{ZHOU2016TowardAutomaticModelComparison} argued that rCESS was more appropriate. Here we confirm that this is also the case in a phylogenetic context. We provide two experiments. In the first experiment, we simulated one unrooted tree of $10$ taxa, and  generated one data set of DNA sequences and each sequence has length $100$. The setup of tree simulation is the same as Section \emph{Simulation Studies}. We ran adaptive annealed SMC algorithm in two schemes: (a) $rCESS_{r} = 0.99$; (b) $rESS_{r} = 0.99$. The one based on $rESS_{r}$ only differs in the way NextAnnealingParameter is implemented; shown in Algorithm~\ref{algo:NextannealingparameterESS}. We used $K = 1000$ particles. Resampling of particles was triggered when $rESS < 0.5$. Figure \ref{fig:CESS and ESS} demonstrates the 
advantage of using rCESS over rESS in adaptive annealed SMC. The annealing parameter difference $(\phi_{r} - \phi_{r-1})$ increases smoothly in the rCESS scheme, while in the rESS scheme there are big gaps in annealing parameter increment after doing resampling, then the consecutive annealing parameter change decreases gradually until the next resampling time. The number of iterations $R$ for adaptive annealed SMC using rESS is much larger than using rCESS. 

\begin{algorithm}
	\caption{\bf{Alternative NextAnnealingParameter procedure (sub-optimal) }}
	\label{algo:NextannealingparameterESS}
	{\fontsize{12pt}{12pt}\selectfont
		\begin{algorithmic}[1]
			\State {\bfseries Inputs:} (a) Particle population from previous SMC iteration $(x_{r-1,\cdot}, w_{r-1,\cdot})$;
			(b) Annealing parameter $\phi_{r-1}$ of previous SMC iteration;
			(c) A degeneracy decay target $\alpha \in (0, 1)$.  
			
			\State {\bfseries Outputs:} automatic choice of annealing parameter $\phi_r$.
			
			\State  Initialize the function $\tilde g$ assessing the particle population quality associated to a putative annealing parameter $\phi$:
			\[ \tilde g(\phi) = \left( 
			\sum_{k=1}^K W_{r-1,k} p(y|x_{r-1,k})^{\phi - \phi_{r-1}} 
			\right)^2 
			\Big\slash
			\sum_{k=1}^K (W_{r-1,k} 
			p(y|x_{r-1,k})^{\phi - \phi_{r-1}})^2,\] 
			\If{$\tilde g(1) \ge \alpha \tilde g(\phi_{r-1})$}
			\State return $\phi_r = 1$. 
			\Else
			\State return $\phi_r = \phi^* \in (\phi_{r-1}, 1)$ such that $\tilde g(\phi^*) = \alpha \tilde g(\phi_{r-1})$ via bisection. 
			\EndIf		
		\end{algorithmic}
	}
\end{algorithm}

\begin{figure}
    \centering
        \includegraphics[width=0.9\textwidth]{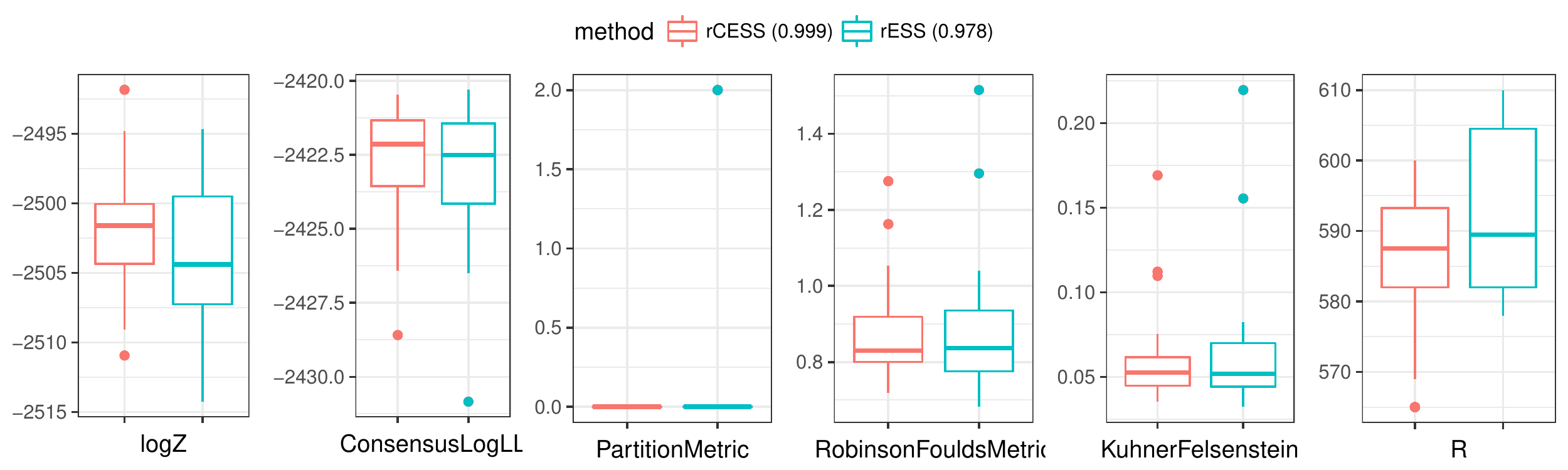}
        \caption{Comparison of adaptive annealed SMC using rCESS and rESS in terms of estimating the log marginal likelihood, the   log likelihood of the consensus tree, tree distance metrics, and the number of SMC iterations ($R$).}
        \label{fig:ESSvsCESS15}
\end{figure}  

In our second experiment, we compared the performance of adaptive annealed SMC using rCESS and rESS in terms of tree metrics and marginal likelihood. We simulated one unrooted tree of $15$ taxa, and generated one data set of DNA sequences. Each sequence has length $200$. The tree simulation setup is the same as Section \emph{Simulation Studies}. We ran adaptive annealed SMC algorithm $20$ times with $rCESS_{r} = 0.999$ and $rESS_{r} = 0.978$ respectively. The number of particles was set to $K = 500$. Under this setting, the computational costs of the two schemes are quite similar.  The numbers of annealing parameters ($R$) selected via rCESS and rESS are similar. Figure \ref{fig:ESSvsCESS15} displays the marginal likelihood estimates, consensus likelihood and tree metrics provided by adaptive SMC using rCESS and rESS. The log  marginal likelihood estimates and log consensus likelihoods provided by adaptive SMC using rCESS are higher and have lower variability. The PF, RF and KF metrics provided by the two schemes are quite close, while the metrics provided by rCESS scheme have lower variability.

\section{APPENDIX 6} 

\subsection{Estimates of Marginal Likelihood from LIS}

We described the LIS procedure as follows:
\begin{enumerate}
\item Sample an index $v_{0}$ randomly from $\{1, 2, \ldots, N\}$, and sample $x_{0, v_{1}}\sim \pi_{0}(\cdot)$.
\item For $d = 0, 1, \ldots, D$, sample $N$ states from $\pi_{d}$ as follows:
\begin{enumerate}
\item If $d>0$: sample an index $v_{d}$ from $\{1, 2, \ldots, N \}$, and set $x_{d,v_{d}} = x_{d-1*d}$.
\item For $k = v_{d}+1, \ldots, N$, sample $x_{d,k}$ from the forward kernel $x_{d,k}\sim K_{d}(x_{d,k-1}, \cdot)$.
\item For $k = v_{d}-1, \ldots, 1$, sample $x_{d,k}$ from the backward kernel $x_{d,k}\sim L_{d}(x_{d,k+1}, \cdot)$.
\item If $d <D$, sample $\mu_{d}$ from $\{1, 2, \ldots, N_{d} \}$ according to the following probabilities:
\[
p(\mu_{d}|x_{d}) = \frac{\gamma_{d-1*d}(x_{d,\mu_{d}})}{\gamma_{d}(x_{d,\mu_{d}})}\bigg/\sum_{k=1}^{N_{d}}\frac{\gamma_{d-1*d}(x_{d,k})}{\gamma_{d}(x_{d,k})},
\]
and set $x_{d*d+1}$ to $x_{d,\mu_{d}}$.
\end{enumerate}
\item Compute the likelihood estimate 
\[
\hat{Z}_{LIS} = \prod_{d = 1}^{D}\bigg[\frac{1}{N}\sum_{k=1}^{N}\frac{\gamma_{d-1*d}(x_{d-1,k})}{\gamma_{d-1}(x_{d-1,k})}\bigg/\frac{1}{N}\sum_{k=1}^{N}\frac{\gamma_{d-1*d}(x_{d,k})}{\gamma_{d}(x_{d,k})} \bigg].
\]
\end{enumerate}
Note that if the backward kernel is reversible, then the forward kernel is the same as backward kernel. In this paper, we use the MCMC kernel as backward and forward kernels in LIS.

\section{APPENDIX 7} 

\subsection{Derivation of Upper Bound of CV}

\begin{eqnarray*}
CV &=& \frac{sd(\hat{Z})}{\E(\hat{Z})} \\
&=& \frac{\sqrt{\frac{1}{n}\sum_{i = 1}^{n}(\hat{Z}_{i} - \frac{1}{n}\sum_{i = 1}^{n}\hat{Z}_{i})^{2}}}{\frac{1}{n}\sum_{i = 1}^{n}\hat{Z}_{i}} \\
&=& \sqrt{n}\sqrt{\sum_{i=1}^{n}\bigg(\frac{\hat{Z}_{i}}{\sum_{i = 1}^{n}\hat{Z}_{i}} -\frac{1}{n}\bigg)^{2}}
\end{eqnarray*}
For non-negative $Z_{i}$, the CV is maximized when $\frac{\hat{Z}_{i}}{\sum_{i = 1}^{n}\hat{Z}_{i}} = 1$ for some $i$, and $0$ for the rest. In this extreme case, $\hat{Z}_{i}$ is much larger than the rest. The upper bound of the CV can be simplified to $\sqrt{n-1}$.

\section{APPENDIX 8} 

\subsection{Tuning of $\beta$ and $K$}

\begin{figure}
\includegraphics[width=1\textwidth]{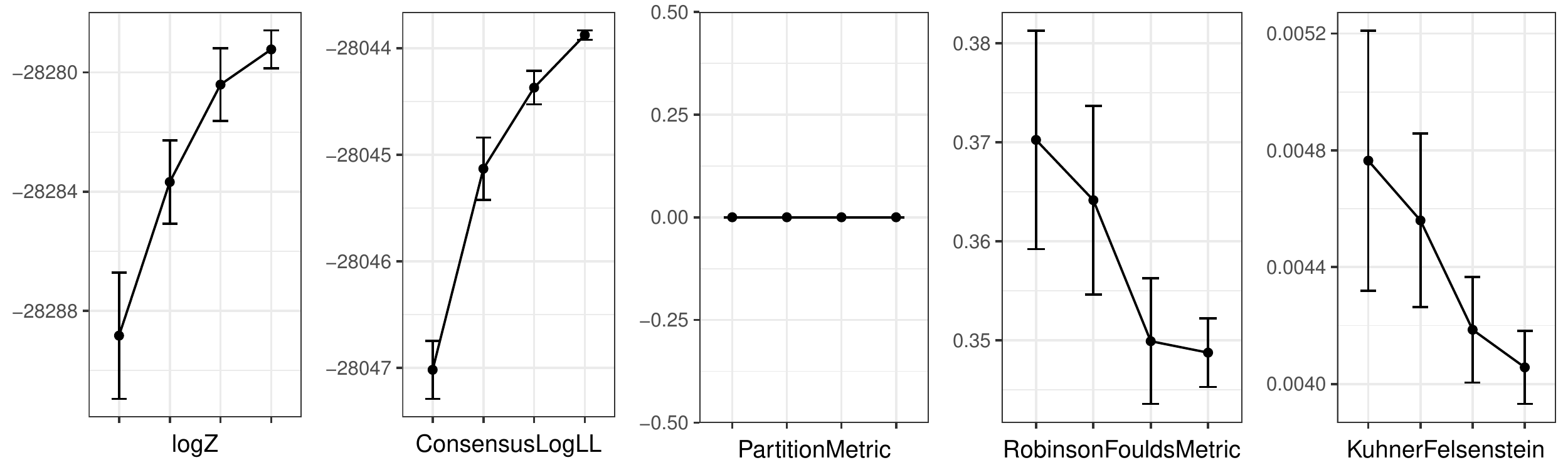}
\caption{Comparison of adaptive SMC algorithm with different numbers of particles, from left to right $K = 100, 300, 1000, 3000$.}
\label{fig:Difparticles}
\end{figure}

\begin{figure}
\includegraphics[width=1\textwidth]{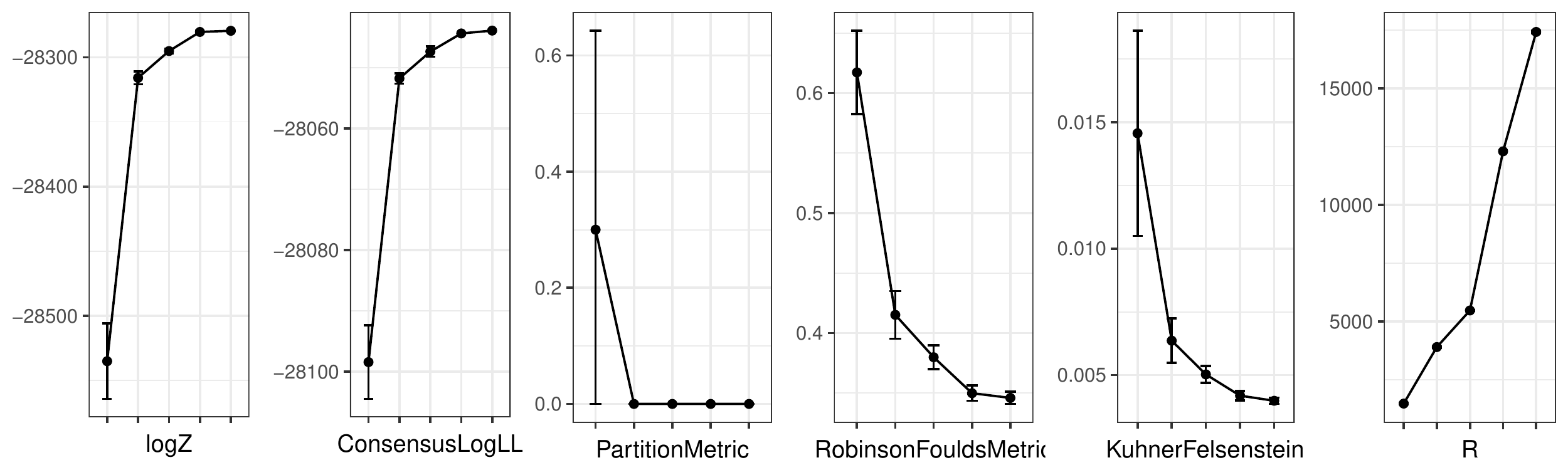}
\caption{Comparison of adaptive SMC algorithm with different $\beta$, from left to right $\beta = 3, 4, 4.3, 5, 5.3$. Here $R$ is the total number of SMC iterations.}
\label{fig:SMCalpha}
\end{figure}

In Figure \ref{fig:Difparticles}, we compare the performance of ASMC algorithm as a function of $K$, with $\beta$ fixed at $5$. We used four different numbers of particles $K = 100, 300, 1000, 3000$. Both the marginal likelihood estimate and tree metrics improve as $K$ increases. Figure \ref{fig:SMCalpha} displays the performance of ASMC algorithm as a function of $\beta$, with $K = 1000$. 
 $R$ is the total number of SMC iterations.
We used five distinct $\beta$ values, $\beta = 3, 4, 4.3, 5, 5.3$.
The marginal likelihood estimates and tree metrics improve as $\beta$ increases; they tend to be stable after $\beta$ reaches $5$. A larger value of $\beta$ can improve the performance of ASMC more significantly than an increase in $K$. 

\section{Appendix 9}

\subsection{Comparison of ASMC, DASMC, LIS and SS for large $K$}

In this experiment, we focus on evaluating the marginal likelihood estimates using ASMC, DASMC, LIS and SS with a shared, large computational budget.  
We simulated an unrooted tree of $4$ taxa, generated one data set of DNA sequences of length $10$. Every algorithm for each data set was repeated $50$ times with different random seeds. We set $\beta = 2$ and $K=200000$. 
The setup of DASMC, LIS and SS is the same as Section \emph{Comparison of marginal likelihood Estimates}.

\begin{figure}
    \centering
        \includegraphics[width=0.5\textwidth]{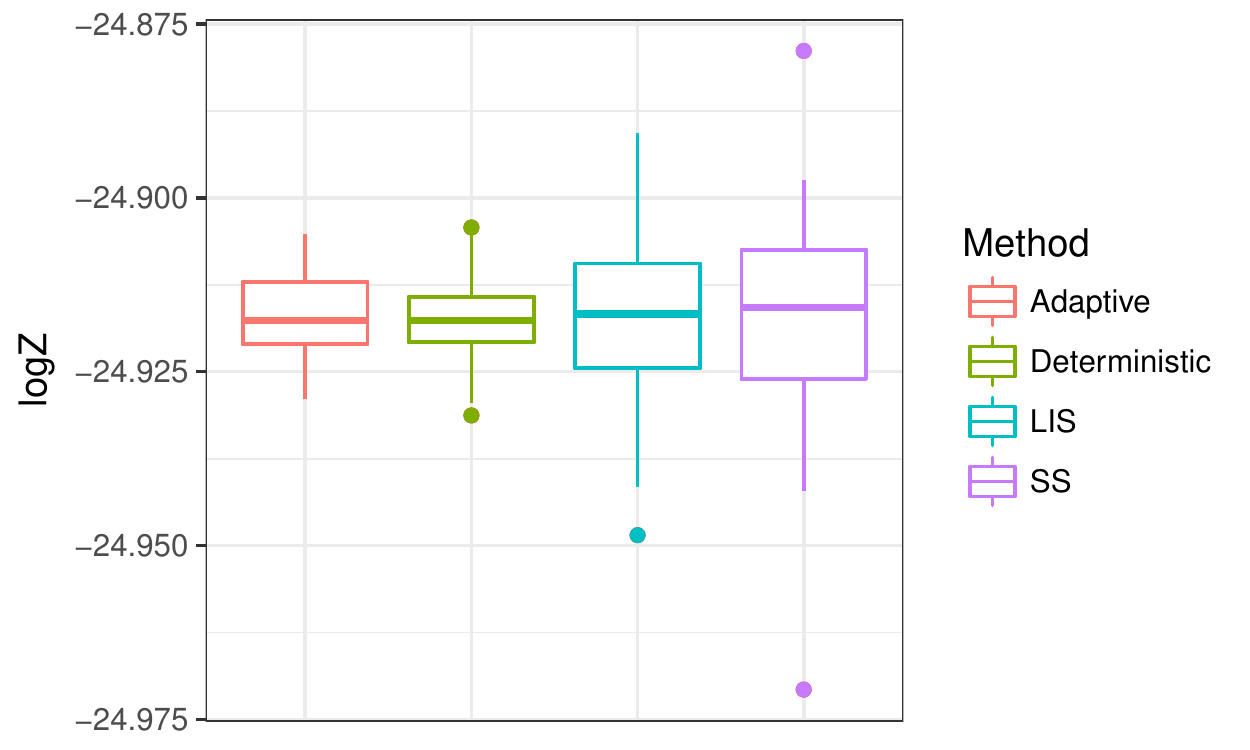}
        \caption{Comparison of marginal likelihood (in log scale) provided by ASMC, DASMC, LIS and SS with a fixed computational budget when $K = 200000$. }
        \label{fig:logZlimiting}
\end{figure}

Figure \ref{fig:logZlimiting} shows the comparison of the performance of the four algorithms in terms of the marginal likelihoods in the log scale. 
The mean log marginalized likelihood estimates provided by ASMC, DASMC, LIS and SS are quite close. The variance of estimates for ASMC and DASMC is smaller than LIS and SS. 

\end{document}